\def\vec#1{{\rm\bf #1}}
\documentclass[fleqn]{vch-book}
\usepackage{amsmath}
\usepackage{amssymb}
\usepackage{makeidx}
    \makeindex
\usepackage{times}
\usepackage{tabularx}

\usepackage{epsfig}

\begin{document}

\author{Hern\'an A. Makse$^1$, Jasna Bruji\'c$^2$, and Sam F. Edwards$^2$\\
\\
$^1$ Levich Institute and Physics Department, \\
City College of New York,
New York, NY 10031, US \\
$^2$ Polymers and Colloids Group, Cavendish Laboratory, \\University of
Cambridge, Madingley Road, Cambridge CB3 0HE, UK
}

\title{Statistical Mechanics of Jammed Matter}

\maketitle
\setcounter{page}{4}

\begin{preface}
A thermodynamic formulation of jammed matter is reviewed.
Experiments and simulations of compressed emulsions and granular
materials are then used to provide a foundation for  the
thermodynamics.

\vspace{1cm} Acknowledgments: The authors would like to thank
R. Ball, R. Blumenfeld, D. Bruji\'c, D. Grinev, H. Herrmann,
 I. Hopkinson, J. Kurchan,
and M. Shattuck
for stimulating discussions
and F. Potiguar, C. Song, P. Wang and  H. Zhang for critical
reading of the manuscript.
H. A. Makse is grateful to the
hospitality of the Isaac Newton Institute at the University of
Cambridge where this work was done. H. A. Makse acknowledges
financial support from the Department of Energy, Division of Basic
Science, Division of Materials Sciences and Engineering, DE-FE02-03ER46089,
and the National Science Foundation, DMR Materials Science
Program, DMR-0239504.

\end{preface}

\tableofcontents

\chapter{Introduction to the Concept of Jamming}
\label{intro}
%
%
%
%
%
%
%
%



The act of jamming or the condition of being jammed stems from the
following layman meaning:
\begin{description}
    \item [`a crowd or congestion of people or things in a limited space,
e.g. a traffic jam'.]
\end{description}
The scientific translation defines ``jamming'' as a state which
emerges when a many-body system is blocked in a configuration far
from equilibrium, from which it takes too long a time to relax for
the time scale to be a measurable quantity. Jamming is emerging as
a fundamental feature of many diverse systems \cite{liu1}, such as

\begin{itemize}
    \item Granular materials: sand, sugar, marbles, dry powders
    \item Emulsions: mayonnaise, custard, milk
    \item Colloidal suspensions: paints, muds
    \item Structural glasses: polymer melts, silica glass
    \item Spin glasses.
\end{itemize}

These distinct disordered systems are but a few examples of out of
equilibrium systems, which are united by their behaviour at the
point of structural arrest. Whereas one can think of liquids or
suspensions as consisting of particles which move very slowly
compared to gases, a state may occur where all particles are in
close contact with one another and therefore experience jamming.
The process of jamming is specific to the system in question due
to their different microscopic properties. The following examples
illustrate this. While it suffices to pour a granular material
into a closed container and shake it to jam up the particles, the
emulsion droplets require a large `squeezing' force usually
implemented by centrifugation. On the other hand, the interaction
between colloidal particles can be tuned such that the
interparticle attraction induces a jammed configuration even at
low densities of the material. Furthermore, glassy materials can
be cooled down to very low temperatures at which the molecules can
no longer diffuse, thus trapping the system into jammed
configurations. Hence, through very different jamming mechanisms,
we arrive at the jammed state for a variety of systems.

All these systems belong to a class of materials known as `soft'
matter, referring to their complex mechanical properties which are
neither fluid nor solid-like. This behaviour is directly linked to
the material's capability to support a mechanical disturbance once
it has reached a jammed state. While the concepts of crowding and
the subsequent mechanical response unite these materials, the
details of their constitutive particles introduce important
differences. For instance, in a polymer melt, it is the physical
chemistry of the individual strands which will govern the
ensemble, whereas it is the interactions between the colloidal
particles, rather than their constituent molecules, which will
determine the system behaviour in suspensions. Moreover, particles
of sizes up to $1\mu$m
 are governed by the laws of statistical mechanics since their
dynamics is due to thermal (Brownian) motion. Above that threshold
size (e.g. grains), the gravitational energy  exceeds $k_B T$,
where $T$ is the room temperature and $k_B$ the Boltzmann
constant, thus prohibiting motion. The colloidal regime is
therefore defined for sizes between $1$nm and $1\mu$m, such that
thermal averaging is present. It applies to glasses, colloids,
surfactants and microemulsions, or in other words, to `complex
fluids'. Most of the fundamental physics research has been
performed on thermal systems until present, and many of the
unifying concepts have arisen through the comparison of systems
within this category.

In the next two sections we first describe the structural arrest
in thermal systems, where the classical statistical mechanics
tools are applicable, and then proceed to athermal systems, such
as granular materials and compressed emulsions,
 in which new situations suitable for a statistical
analysis are introduced.

\section{Jamming in Glassy Systems}

\begin{figure}
         \begin{center}
          \epsfig{file=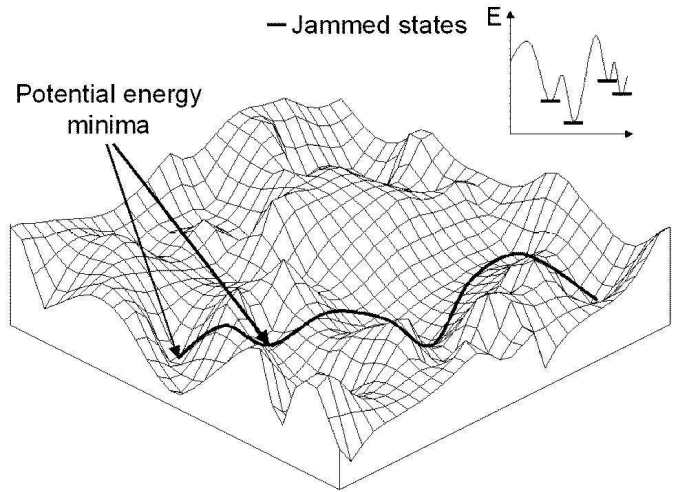,width=9 cm}
         \end{center}
\caption{The multidimensional energy landscape dominates the
dynamics below the glass transition, as the system explores the
inherent structures defined as the potential energy minima. A
trajectory through the landscape is shown and the analogy between
these inherent structures and the jammed states in granular
materials is examined.} \label{landscape}
\end{figure}

In a fluid at thermal equilibrium the particle dynamics is too
fast to capture the detail of the underlying potential energy
landscape, thus it appears flat. Decreasing the temperature slows
down the Brownian dynamics, implying a limiting temperature below
which the system can no longer be equilibrated in this way. Hence,
the thermal system falls out of equilibrium on the time scale of
the experiment and thus undergoes a {\it glass transition}
\cite{debenedetti}. The motion of each particle is no longer
thermally activated and only the vibration inside the cage formed
by its surrounding neighbours persists. However, even below the
glass transition temperature the particles continue to relax, but
the nature of the relaxation is very different to that in
equilibrium. This phenomenon of a structural evolution beyond the
glassy state is known as ``aging''. The dynamics becomes dominated
by the multidimensional potential energy surface which the system
can explore as a function of the degrees of freedom of the
particles, depicted in Fig. \ref{landscape}. In order to describe
this landscape Stillinger and coworkers \cite{stillinger}, based
on ideas introduced  by Goldstein \cite{goldstein}, developed the
concept of inherent structures which are defined as the potential
energy minima. The trajectory of a system aging at temperature $T$
can be mapped onto the successive potential basins that the system
explores. Computational methods are the only available technique
for investigating this behaviour, in which the inherent structures
are found via steepest-descent quenching of the system
configurations to the basins of the wells. The entropy of the
system can be shown to be separable into contributions from the
available configurations and the vibrational modes around each
minimum. There have been many studies which have embarked on an
investigation of aging through the exploration of the
configurational space \cite{sastry,kob,parisi}.

The importance of the inherent structure formalism is in enabling
the comparison of jamming in particulate systems with glasses
\cite{nicodemi-is}. The entropy arising from the inherent
configurations of the glass at very low temperatures and the
exploration of these configurations due to the vibrational modes
of the particles could be viewed as analogous to the
configurational changes in particulate packings under slow tapping
or shear. However, in granular materials there is an added effect
of friction, which dissipates the analogous vibrations at once.
Unlike granular materials, a thermal system is never permanently
trapped in the bottom of a valley, but escapes in other accessible
unstable directions through intrinsic thermal vibrations. At any
finite temperature the system will not resemble the granular
system in that it continuously evolves toward a maximum density
state. Thus, the only true analogous situation between glasses and
granular materials is valid at zero temperature. However, there
are characteristic features of the glassy relaxation at a finite
$T$ which act as useful tools for the description of granular
systems by exploiting the analogy between the relaxation of
powders  and aging in glassy systems \cite{struik}.

For instance, theories developed during the late eighties and
nineties in the field of spin glasses \cite{virasoro,ktw}
have led to a better understanding of glassy systems through the
generalisation of usual equilibrium relations, such as the
fluctuation-dissipation relation, to situations far from
equilibrium \cite{ck}. This approach developed by Cugliandolo,
Kurchan and collaborators yielded macroscopic observable
properties, such as an ``effective temperature'' for the slow
modes of relaxation, which could then be compared between various
glassy systems. Furthermore, the existence of an effective
temperature with a thermodynamic meaning in glasses at very low
temperature suggests an `ergodicity' for the long-time behaviour
of the system \cite{kurchan1}. This ergodicity is closely related
to the statistical ideas for granular systems
\cite{kurchan2,nicodemi,bklm} which we will introduce in the
following sections. In support of this argument, the effective
temperature in glasses is found to be an adequate concept for
describing granular matter \cite{nature}, as it will be shown in
Section \ref{simulations}.



From a theoretical point of view, these systems are still only
understood in terms of predictions of a general nature and many
open questions remain. There is still much debate on issues such
as the precise mechanisms of surfing the energy landscape, the
effects of memory in the system, the slowing down of the system
with time, and the discrepancies between the behaviour of
different glassy systems, but they are beyond the scope of this
work.

\section{Jamming in Particulate Systems}
\label{granularjamming}

In a sense one would imagine there is no simpler physical system
than a granular assembly. After all it is just a set of packed
rigid objects with no interaction energy. It is the inability to
describe the system on the continuum level in any other way except
according to its geometry which has led to a lack of a well
established granular theory until present. Mostly due to their
industrial importance, there has been a vast literature describing
phenomenological observations without an encompassing theory. In
the words of de Gennes, the state of granular matter can be
compared to solid state physics in the 30's or critical phenomena
in phase transitions before the renormalization group. In other
words, there is a need for describing the universal features of
the observed behaviour within a theoretical framework devised for
these and other jammed systems.

In parallel with the extensive research on glasses, described
earlier, a decade ago Edwards and collaborators postulated the
existence of a statistical ensemble for granular matter, despite
the lack of thermal motion and the absence of an equilibrium state
\cite{edwards1,edwards2,edwards3,edwards4,edwards5}. The main
postulate was based on jamming the granular particles at a fixed
total volume such that all microscopic jammed states are equally
probable and become accessible to one another (ergodic hypothesis)
by the application of a type of external perturbation such as
tapping or shear, just as thermal systems explore their energy
landscape through Brownian motion. Hence, let us consider granular
jamming in more detail.

Pouring sugar into a cup is the simplest example of a fluid to
solid transition which takes place solely because of a density
increase.
In terms of physics, in particulate materials such as emulsions
and granular media, a jammed system results if particles are
packed together so that all particles are touching their
neighbours, which obviously requires a sufficiently high density.
In these athermal systems there is no kinetic energy of
consequence; the typical energy required to change the positions
of the jammed particles is very large compared to the thermal
energy at room temperature ($\sim 10^{14}$ times, see Section
\ref{simulations}). As a result, the material remains arrested in
a static state and is able to withstand a sufficiently small
applied stress.

There is a subtle, but crucial difference, between a configuration
in mechanical equilibrium and a jammed configuration, particularly
in the context of this research. The mechanism of arriving at a
{\it static configuration} by an increase in density, which is an
intuitively obvious process, is not always sufficient to satisfy
the {\it jamming} condition in our definition. This applies
especially to systems which bear knowledge of the process of their
creation. For instance, pouring grains into a container results in
a pile at a given angle of repose. This mechanical equilibrium
configuration is not jammed because in response to an external
perturbation, the constituent particles will irreversibly
rearrange, approaching a truly jammed configuration. The
statistical mechanics which we are aiming to test implies an
ergodic hypothesis, which is not valid in such history-dependent
samples \footnote{Theories attempting to describe such systems
have been developed by Bouchaud {\it et al.}, proposing a model
for the `fragile' systems, i.e. systems which rearrange under
infinitesimal stresses \cite{bouchaud}. However, this situation
will not be considered here.}.

It turns out that by allowing the system to explore its available
configurational space through external mechanical perturbations,
the system will rearrange such that all possible configurations
(w.r.t. the perturbation) become accessible to one another.
Continuing with the analogy in the real world, the gentle tapping
on a table of the cup filled with sugar will initially change the
unstable angle of repose of the sugar pile and flatten its top
surface, and therefore its density, until it settles into a
desired configuration which depends on the strength of the tap.
We can only perform a statistical analysis on the resulting
configurations which have no memory of their creation, i.e. the
true jammed configurations. Thus we arrive at a jammed ensemble,
suitable for the application of statistical mechanics, described
in Section \ref{theory}. Since the particles can jump across the
energy landscape during the tap, but then stop at once due to
frictional dissipation, there is an analogy to the inherent
structure formalism in glassy systems. This new statistical
mechanics is able to provide unifying concepts between previously
unrelated media.

\subsection{Applications of the jamming condition}

The statistical mechanics which we are aiming to develop implies
an ergodic hypothesis, which is not valid in history-dependent
samples. In fact, there are many experimental situations in which
the statistical mechanics cannot be applied due to the lack of
ergodicity. For instance convection cycles have been observed in
granular systems under vigorous tapping \cite{knight}- an effect
which is closely associated with the segregation process of
different granular species. These types of closed loops in phase
space cannot be described within the thermodynamic framework.
Rapid granular flows observed in pouring sand in a pile, or
vigorously shaken granular systems at low density are out of the
scope of the present approach since the systems are exploring
configurations far from the jammed states \cite{savage}. Kinetic
theories of inelastic gases are more appropriate to treat these
situations \cite{jenkins}. The physics of the angle of repose
\cite{herrmann} may not be understood under the thermodynamic
framework due to the absence of the jamming condition of the pile,
despite the fact that it is static. In many practical situations,
heterogeneities appear which also preclude the application of a
thermodynamic approach. For instance, when granular materials are
sheared in a sufficiently large shear stage, shear bands appear
where the strain is discontinuous \cite{powders}. Such local
effects cannot be captured by the present thermodynamic approach.

On the other hand, if the application of statistical physics to
jammed phenomena were to prove productive, then one could
anticipate a more profound insight into the characterisation and
understanding of the system as a whole. For instance,  the
thermodynamic hypothesis would lead to the prediction of
macroscopic quantities such as viscosity and complex shear moduli,
which would in turn provide a complete rheological caracterisation
of the system. As a matter of theoretical interest, a statistical
ensemble for jammed matter could be one of the very few
generalisations of the statistical mechanics of Gibbs and
Boltzmann to systems out of equilibrium.

\subsection{Achieving the jammed state}
\label{jammedstate}

\begin{figure}[tbp]
         \begin{center}
          \epsfig{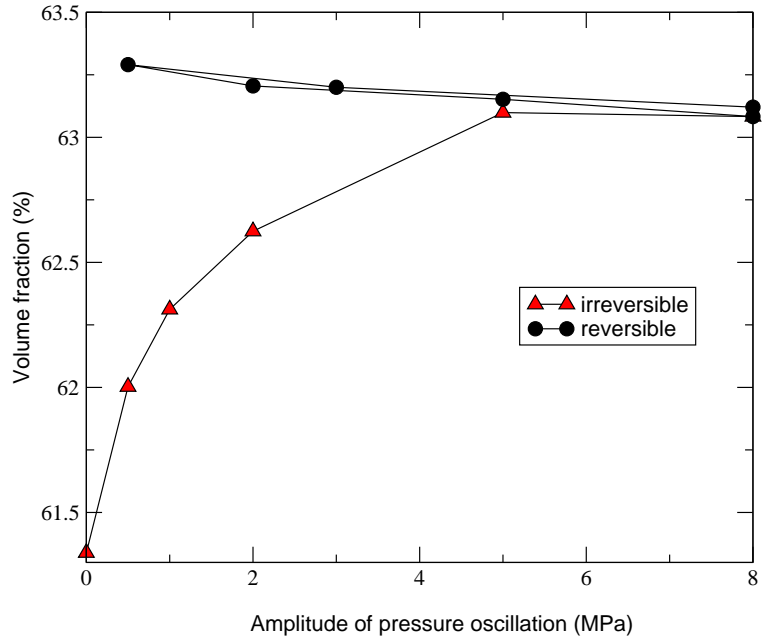}
         \end{center}
\caption{Compaction curve for a packing of glass beads under an
oscillating pressure. Increasing the amplitude of oscillation
initially increases the density by filling the loose voids, after
which a reversible regime is achieved (from \protect\cite{sdr}).}
\label{compaction}
\end{figure}


Experimentally, the conditions for a statistical ensemble of
jammed states can be achieved by pre-treating the granular
assembly by tapping or via slow shear-driving. Experiments at the
University of Chicago involving the tapping of granular columns
were the first to show the existence of a reversible regime in
which the system configurations are independently sampled
\cite{chicago1}. Starting with a loose packing of the grains, the
tapping routine initially removes the unstable loose voids and
thus eliminates the irreversible grain motion. Once all the grains
are touching their neighbours, the density of the resulting
configuration becomes dependent on the tapping amplitude and the
number of taps; the larger the amplitude, the lower the density.
The mechanism of the {\it compaction} process leading to a
steady-state density is extremely slow, in fact, it is logarithmic
in the number of taps. This dependence of the density of grains on
the external perturbation of the system once the memory effects of
the pile construction details have been removed, is known as the
reversible branch of the `compaction curve', see Fig.
\ref{compaction}. Despite the presence of friction between grains
(implying memory effects) this curve is reversible, establishing a
new type of equilibrium states. It is along this curve that the
thermodynamics for granular matter can be applied.

There have been several further experiments confirming these
results for different system geometries, particle elasticities and
compaction techniques. For example, the system can be mechanically
tapped or oscillated, vibrated using a loudspeaker, slowly sheared
in a couette geometry, or even allowed to relax under large
pressures over long periods of time, all to the same effect
\cite{chicago2,bideau,sdr,cavendish}. Here we show a new
compaction regime under an oscillating pressure where the same
density dependence of a packing of glass and acrylic beads is
noted for varying amplitudes of the pressure oscillation. These
experiments have been performed at Schlumberger-Doll Research
\cite{sdr}. The resulting curve of the achieved volume fraction as
a function of the amplitude of the pressure oscillation is shown
in Fig. \ref{compaction}.

Moreover, experiments in the Cavendish laboratory \cite{cavendish}
have shown how the conductivity of powdered graphite can also be a
measure of the particle density as it is being vibrated, in which
the direct link to the volume function is less obvious, but the
qualitative results indicate the same trends. The methodology for
achieving jammed configurations has also been established
numerically for the purpose of rheological and thermodynamical
studies and it will be described in Section \ref{simulations}.

\begin{figure}[tbp]
         \begin{center}
          \epsfig{file=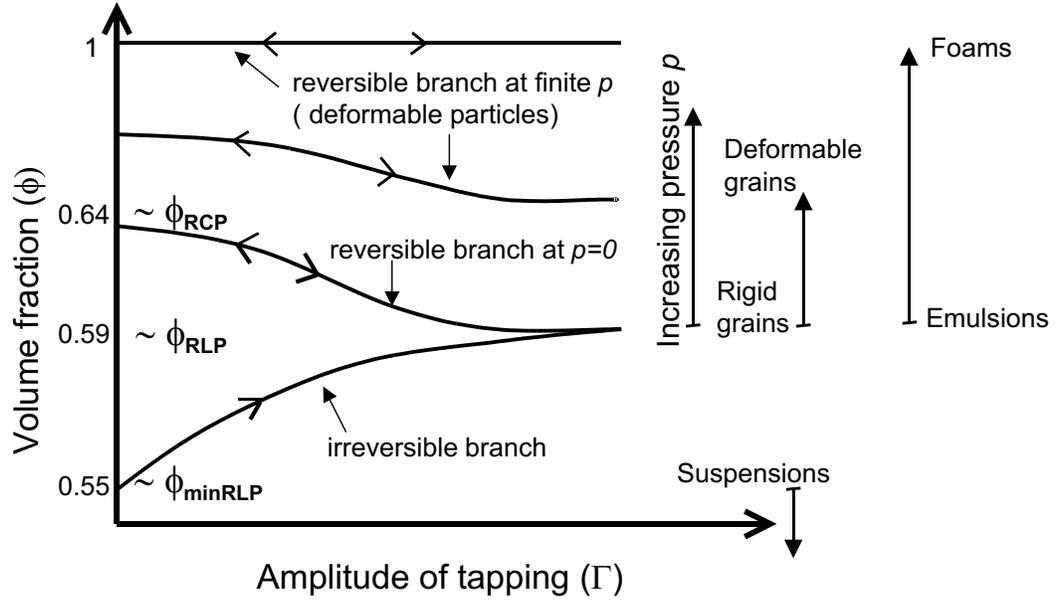,width=14 cm}
         \end{center}
\caption{Compaction curves of volume fraction $\phi$ versus
amplitude of oscillation $\Gamma$ for different external confining
pressures, $p$. Increasing the amplitude of oscillation initially
increases the volume fraction by filling the loose voids
(irreversible branch), after which a reversible regime is
achieved. For infinitely rigid grains (the ``zero pressure''
curve) the minimum volume fraction along the irrevesible branch is
the random loose packing. The reversible branch goes from the
random loose packing fraction to the random close packing
fraction. Below the minimum RLP only suspensions can exist.
} \label{comp}
\end{figure}

At this point, it is important to note that we have only
considered infinitely rigid, rough grains  in which an increase in
the pressure of the system, for instance by placing a piston on
top of the grains, causes no change in the shape of the grains and
therefore no change in the packing density. On the other hand,
real grains have a finite elastic modulus, thus the application of
a sufficiently large external pressure will always result in grain
deformation and therefore a density increase unrelated to the
tapping. In soft particles, such as emulsions, the effect of
pressure is more significant. The tapping experiment described
above measured the resulting densities at atmospheric pressure,
which is considered to be the zero reference pressure. The same
experiment can be repeated at finite pressures giving rise to
equivalent compaction curves, depicted in Fig. \ref{comp}. Whereas
hard grains, such as glass beads, require extremely large
pressures ($\sim 1$MPa) to deform and the amount of deformation is
limited by their yield stress, softer particles, such as rubber,
are able to reach higher densities with relative ease. Droplets
and bubbles, being the softest particles one can have, are capable
of reaching the density of 1, corresponding to a biliquid foam and
a foam, respectively, by an application of much smaller pressures
($\sim 1$kPa). They have the advantage of the whole pressure range
being accessible to them. Another distinction between granular
materials and emulsions is the presence of friction in the former
and the smoothness of the latter. Since friction plays an
important role in inducing memory into the system, its absence
leads to a much easier achievement of the jammed state, described
above. For instance, in the case of emulsions, allowing the
particles to cream under gravity will suffice to arrive at the
reversible part of the compaction curve, bypassing the
irreversible branch as it will be shown in Section
\ref{experiments}.

\section{Unifying Concepts in Granular Matter and Glasses}


In the preceding paragraphs, it has been shown under which
conditions both thermal and athermal systems explore the
configurational energy landscape, which possibly results in
commonalities in their behaviour. At present, new unifying
theoretical descriptions for jammed matter are being sought, as
well as new experimental evidence to unify the predicted state for
{\it all} varieties of jammed systems.

The prediction of how different systems jam with respect to the
applied stress, density and temperature has led to a speculative
diagram proposed by Liu and Nagel in their article ``Jamming is
not just cool anymore'' in Nature \cite{liu-nagel,trappe,anna}. It
links the behaviour of glasses (thermal systems) and bubbles,
grains, droplets (athermal systems) by the dynamics of their
approach to jamming.


Since the observable properties such as applied strain,
temperature and density can be obtained by consideration of only
the jammed configurations in a given system, the thermodynamics of
jamming, discussed in the next section, is intimately related to
the ideas put forward in the jamming phase diagram.


\chapter{New Statistical Mechanics for Granular Matter}
\label{theory}


%
This Section aims to justify the use of statistical mechanics
tools in situations where the system is far from thermal
equilibrium, but jammed. In what follows, we present the classical
statistical mechanics theorems to an extent which facilitates an
understanding of the important concepts for the development of an
analogous granular theory, as well as the assumptions necessary
for the belief in such a parallel approach. Thereafter, we present
a theoretical framework to fully describe the exact specificities
of the granular packing, and the shaking scenario which leads to
the derivation  of the Boltzmann equation for a jammed granular
system.



This kind of an analysis paves the path to macroscopic quantities,
such as  the compactivity, characterising the configurations from
the microstructural information of the packing. It is according to
this theory that the jammed configurations obtained from
experiments and simulations are later characterised.


\section{Classical Statistical Mechanics}
\label{edwards}

In the conventional statistical mechanics of thermal systems, the
different possible configurations, or microstates, of the system
are given by points in the phase space of all positions and
momenta \{$p,q$\} of the constituent particles. The equilibrium
probability density $\rho_{\mbox{\scriptsize eqm}}$ must be a
stationary state of Liouville's equation which implies that
$\rho_{\mbox{\scriptsize eqm}}$
 must be expressed only in
terms of the total energy of the system, $E$
\cite{landau-stat-mech}. The simplest form for a system with
Hamiltonian ${\cal H}(p,q)$ is the microcanonical distribution:
\begin{equation}
\rho_{\mbox{\scriptsize eqm}}(E) = \frac{1}
{\Sigma_{\mbox{\scriptsize eqm}}(E)},
\label{equal}
\end{equation}
for the microstates within the ensemble, ${\cal H}(p,q)=E$, and
zero otherwise. Here,
\begin{equation}
\Sigma_{\mbox{\scriptsize eqm}}(E) = \int \delta \Big(E - {\cal
H}(p,q) \Big) ~~ dp ~ dq,
\end{equation}
is the area of energy surface ${\cal H}(p,q)=E$.

Equation (\ref{equal}) states that all microstates are equally
probable. Assuming that this is the true distribution of the
system implies accepting the ergodic hypothesis, i.e. the
trajectory of the closed system will pass arbitrarily close to any
point in phase space.

It was the remarkable step of Boltzmann to associate this
statistical concept of the number of microstates with the
thermodynamic notion of entropy through his famous formula
\begin{equation}
S_{\mbox{\scriptsize eqm}}(E) = k_B \ln \Sigma_{\mbox{\scriptsize
eqm}}(E).
\end{equation}



Thus, in classical statistical mechanics, the total energy of the
system is sufficient to describe the probability density of
states.
Whereas the study of thermal systems has had the advantage of
available statistical mechanics tools for the exploration of the
phase space, an entirely new statistical method, unrelated to the
temperature, had to be constructed for grains.

\section{Statistical Mechanics for Jammed Matter}

We now  consider a jammed granular system composed of {\it rigid}
grains (deformable particles will be considered in Section
\ref{simulations}). Such a system is analogously described by a
network of contacts between the constituent particles in a fixed
volume, $V$, since there is no relevant energy $E$ in the system.
In the case of granular materials, the analogue of phase space,
the space of microstates of the system, is the space of possible
jammed configurations as a function of the degrees of freedom of
the system $\{\zeta\}$.

It is argued that it is the volume of this system, rather than the
energy, which is the key macroscopic quantity governing the
behaviour of granular matter \cite{edwards1,edwards2,edwards3}. If
we have $N$ grains of specified shape which are assumed to be
infinitely rigid, the system's statistics would be defined by a
function ${\cal W}(\zeta)$, a function which gives the volume of
the system in terms of the specification of the grains.

In this analogy  one replaces the Hamiltonian ${\cal H}(p,q)$ of
the system by the volume function, ${\cal W}(\zeta)$. The average
of ${\cal W}(\zeta)$ over all the jammed configurations determines
the volume $V$ of the system in the same way as the average of the
Hamiltonian determines the average energy $E$ of the system.

\subsection{Definition of the volume function, $\cal W$}
\label{W}

 One of the key questions in this analogy is to
establish the `correct' $\cal W$ function, the statistics of which
is capable of fully describing the system as a whole. The idea is
to partition the volume of the system  into different subsystems
$\alpha$ with volume ${\cal W}^\alpha$, such that the total volume
of a particular configuration is
\begin{equation}
{\cal W}(\zeta) = \sum_\alpha {\cal W}^\alpha.
\end{equation}

It could be that considering the volume of the first coordination
shell of particles around each grain is sufficient; thus, we may
identify the partition $\alpha$ with each grain. However,
particles further away may also play a role in the collective
system response due to enduring contacts, in which case $\cal W$
should encompass further coordination shells. In reality, of
course, the collective nature of the system induces contributions
from grains which are indeed further away from the grain in
question, but the consideration of only its nearest neighbours is
a good starting point for solving the system, and is the way in
which we proceed to describe the $\cal W$ function. The
significance of the appropriate definition of $\cal W$ is best
understood by the consideration of a response to an external
perturbation to the system in terms of analogies with the
Boltzmann equation which we will describe in Section
\ref{granularboltzmann}.

\begin{figure}[tbp]
         \begin{center}
          \epsfig{file=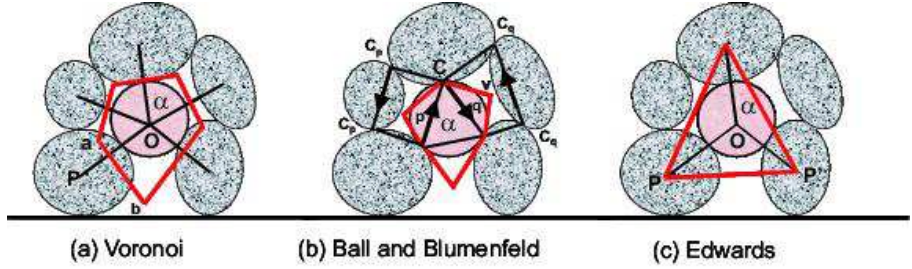,width=12 cm}
         \end{center}
\caption{Different volume functions  as discussed in the text: (a)
Voronoi, (b) Ball and Blumenfeld and (c) Edwards construction.
\label{VOLUMES} }
\end{figure}

Perhaps the most straightforward definition of the function ${\cal
W}^\alpha$ is given in terms of the Voronoi diagram which
partitions the space into a set of regions, associating all grain
centroids in each region to the closest grain centroid, depicted
by line OP in the diagram in Fig. \ref{VOLUMES}a. The loop formed
by the perpendicular bisectors (ab) of each of the lines joining
the central grain to its neighbours is the Voronoi cell, depicted
in red. Even though this construction successfully tiles the
system, its drawback is that there is no analytical formula for
the enclosed volume of each cell. Recently Ball and Blumenfeld
\cite{ball,blumenfeld} have shown by an exact triangulation method
that the volume defining each grain can be given in terms of the
contact points C using vectors constructed from them (see Fig.
\ref{VOLUMES}b). The method consists in defining shortest loops of
grains in contact with one another (p,q loops), thus defining the
void space around the central grain. The difficulty arises in
three dimensions since this construction requires the
identification of void centres, v. This is not an obvious task,
but is currently under consideration. The resulting volume (red)
is the antisymmetric part of the fabric tensor, the significance
of which is its appearance in the calculation of stress
transmission through granular packings \cite{ball}.

A cruder version for the volume per grain, yet with a strong
physical meaning, has been given by Edwards. For a pair of grains
in contact (assumed to be point contacts for rough, rigid grains)
the grains are labelled $\alpha, \beta$,
and the vector from the centre of $\alpha$ to that of $\beta$ is
denoted as $\vec{R}^{\,\alpha\beta}$ and specifies the complete
geometrical information of the packing.
The first step is to construct a configurational tensor $\vec{\cal
C}^{\alpha}$ associated with each grain $\alpha$ based on the
structural information,

\begin{equation}
{\cal C}_{ij}^{\alpha}=\sum_{\beta} R_i^{\alpha \beta}R_j^{\alpha
\beta}. \label{C}
\end{equation}

Then an approximation for the area in 2D or volume in 3D
encompassing the first coordination shell of the grain in question
is given as
\begin{equation} \label{Wfunc}
{\cal W}^\alpha  = 2\sqrt{\mbox{Det} ~{\cal C}_{ij}^\alpha}.
\end{equation}

This volume function is  depicted in the Fig. \ref{VOLUMES}c, with
grain coordination number 3 in two dimensions, where Eq.
(\ref{Wfunc}) should give the area of the triangle (red)
constructed by the centres of grains P which are in contact with
the grain $\alpha$. The above equation is exact if the area is
considered as the determinant of the vector cross product matrix
of the two sides of the triangle, but its validity for higher
coordination numbers and in 3D has not been tested.
 Surprisingly, this approximation works well according to
our experimental studies in  Section \ref{experiments}. This is
due to the partitioning of the obtained volumetric objects into
triangles/pyramids, intrinsic to the method, and subsequently
summing over them to obtain the resulting volume.

This definition is clearly only an approximation of the space
available to each grain since there is an overlap of ${\cal
W}^{\alpha}$ for grains belonging to the same coordination shell.
Thus, it overestimates the total volume of the system: $\sum {\cal
W}^\alpha > V$. However, it is the simplest approximation for the
system based on a single coordination shell of a grain.

\subsection{Entropy and compactivity}
\label{entropycompactivity}

Now that we have explicitly defined $\cal W$ it is possible to
define the entropy of the granular packing. The number of
microstates for a given volume $V$ is measured by the area of the
surface ${\cal W}(\zeta)=V$ in the phase space of jammed
configurations and it is given by:
\begin{equation}
\Sigma_{\mbox{\scriptsize jammed}}(V) = \int \delta\Big(V- {\cal
W}(\zeta) \Big)~~ \Theta(\zeta) ~~ d\zeta,
\end{equation}
where now $d\zeta$ refers to an integral over all possible jammed
configurations and $\delta(V-{\cal W}(\zeta))$ formally imposes
the constraint to the states in the sub-space ${\cal
  W}(\zeta)=V$.
$\Theta(\zeta)$ is a constraint that restricts the summation to
only reversible jammed configurations as opposed to the merely
static equilibrium configurations as previously discussed. This
function will be discussed in detail in Section \ref{ergodicity}.
The radical step is the assumption of equally probable microstates
which leads to an analogous thermodynamic entropy associated with
this statistical quantity:
\begin{equation}
\label{entropy} S(V)=\lambda \ln \Sigma_{\mbox{\scriptsize
jammed}}(V) = \lambda \ln \int \delta\Big(V-{\cal W}(\zeta)\Big)
~~ \Theta(\zeta) ~~ d\zeta,
\end{equation}
which governs the macroscopic behaviour of the system
\cite{edwards3,edwards2}. Here $\lambda$ plays the role of the
Boltzmann constant. The corresponding analogue of temperature,
named the ``compactivity'', is defined as

\begin{equation}\label{compactivity}
X_V^{-1} = \frac{{\partial S} }{{\partial V}}.
\end{equation}
where the subscript $V$ refers to the fact that it is the
derivative of the entropy with respect to the volume.

\begin{figure}[tbp]
         \begin{center}
          \epsfig{file=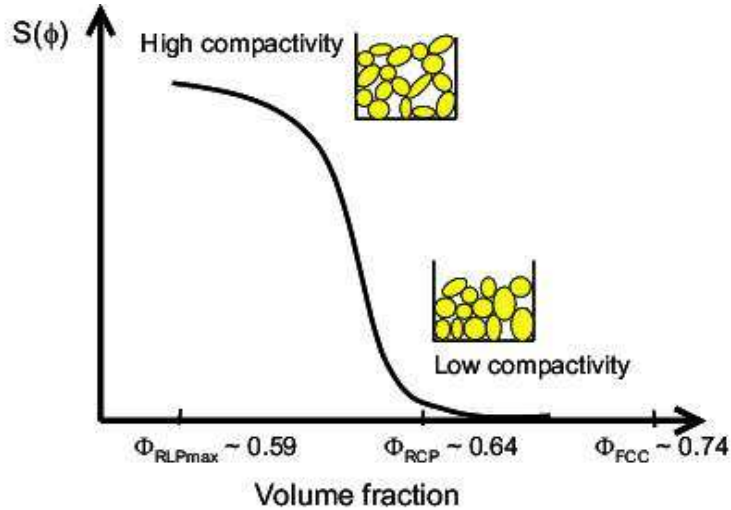,width=10 cm}
         \end{center}
\caption{ Interpretation of the compactivity and entropy in terms
of different packings. See also Fig. \protect\ref{compaction}.
\label{SvsX}}
\end{figure}

This is a bold statement, which perhaps requires further
explanation in terms of the actual role of compactivity in
describing granular systems. We can think of the compactivity as a
measure of how much more compact the system could be, i.e. a large
compactivity implies a loose configuration (e.g. random loose
packing, RLP) while a reduced compactivity implies a more compact
structure (e.g. random close packing, RCP, the densest possible
random packing of monodisperse hard spheres). In terms of the
reversible branch of the compaction curve, large amplitudes
generate packings of high compactivities, while in the limit of
the amplitude going to zero a low compactivity is achieved. In
terms of the entropy, many more configurations are available at
high compactivity, thus the dependence of the entropy on the
volume fraction can be qualitatively described as in Fig.
\ref{SvsX}. In the figure, for monodisperse packings the RCP  is
identified at $\phi\approx 0.64$ \cite{bernal}, the RLP fraction
is identified at $\phi\approx 0.59$ \cite{scott}, while the
crystalline packing, FCC, is at  $\phi = 0.74$ but cannot be
reached by tapping.

At any given tapping amplitude, there exists an equilibrium volume
fraction toward which the system slowly evolves. For instance,  a
system may find itself at a lower entropy than the equilibrium
curve by the application of an internal constraint at a given
volume fraction. This situation can be achieved by creating small
crystalline regions within a packing configuration of a lower
density, and looser regions compensating for the volume reduction
such that the total volume of the system remains constant. This
configuration, given that it is not jammed, will tend toward the
equilibrium packing via the application of a small perturbation by
increasing its entropy. Such an example will be made more explicit
in the derivation of the Boltzmann equation for granular
materials. At volume fractions beyond the RCP (and at atmospheric
pressure) the system is not able to explore the configurations as
they can only be achieved by the partial crystallisation of the
sample, where there are very few configurations available.

It becomes clear from Eq. (\ref{compactivity}) that the
compactivity is only applicable in equilibrated jammed states.  As
an analogue of temperature, it should also obey the zero-th law of
thermodynamics. Hence, two different powders in physical contact
with one another should equilibrate at the same compactivity,
given a mechanism of momentum transfer between the two systems.
Indeed, we may think of an appropriate laboratory experiment which
would test this hypothesis under certain conditions necessary for
creating the analogous situation to heat flow.

\begin{figure}[tbp]
         \begin{center}
          \epsfig{file=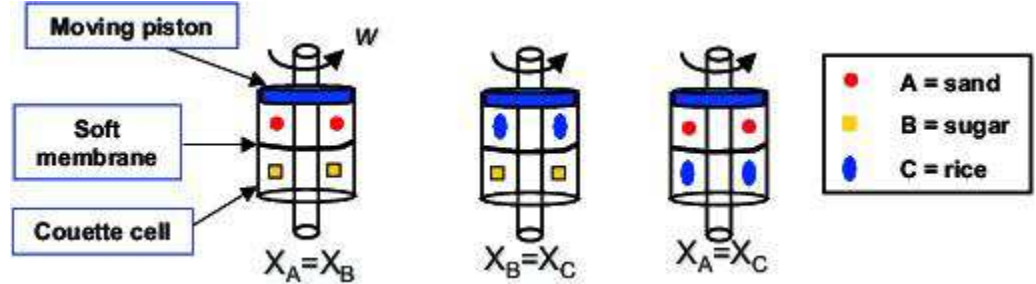,width=14 cm}
         \end{center}
\caption{ABC experiment to test the zero-th law of granular
thermodynamics. \label{abc}}
\end{figure}

Two powders, A and B, of different grain types are poured into a
vertical couette cell as shown in Fig. \ref{abc}. The grains must
experience an equivalent tapping or shearing regime, which is
achieved by the rotation of the inner cylinder of the couette
cell. The species are separated by a flexible diaphragm, such that
momentum transfer between the two systems is ensured. The two
powders must be well separated such that there is no mixing
involved, but in contact nevertheless. The grains are kept at a
constant pressure by a piston which is allowed to move freely to
accommodate for the changes in volume experienced by the two types
of grains. Gravity may play a role in the experiment, which is
avoided by density matching the particles with a suspending fluid.

The experiment consists in placing powders A and B together in the
above cell and slowly shearing them at a given velocity. The
powders should come to equilibrium volumes $V_A$ and $V_B$, with
equivalent respective compactivities, $X_A = X_B= X$. While it is
easy to measure the volumes of the two systems, the measurement of
their compactivities employs more sophisticated methods, discussed
in Section \ref{measuringX}. In the absence of a compactivity
scale, we use powder B as a `thermometer' by placing it in contact
with a third powder C. The volume B is kept at $V_B$ and the
volume of C is allowed to fluctuate until it reaches the
equilibrium state. Finally, powders A and C are put together to
test if they will reach the same volumes as they did in previous
runs in contact with B, thus proving the zero-th law. A form of
the zero-th law of thermodynamics will be shown to be valid
numerically in Section \ref{simulations}.

\subsection{Remarks}
To summarise, the granular thermodynamics is based on two
postulates:

 1) While in the Gibbs construction one assumes that
the physical quantities are obtained as an average over all
possible configurations at a given energy, the granular ensemble
consists of only the jammed configurations at the appropriate
volume.

2) As in the microcanonical equilibrium ensemble, the strong
ergodic hypothesis is that all jammed configurations of a given
volume can be taken to have equal statistical probabilities.

The ergodic hypothesis for granular matter was treated with
skepticism,
mainly because a real powder bears knowledge of its formation and
the experiments are therefore history dependent. Thus, any problem
in soil mechanics or even a controlled pouring of a sand pile does
not satisfy the condition of all jammed states being accessible to
one another as ergodicity has not been achieved, and the
thermodynamic picture is therefore not valid. This point has been
discussed in Section \ref{granularjamming}.
The Chicago experiments of tapping columns \cite{chicago2} showed
the existence of reversible situations. For instance, let the
volume of  the column be $V(n,\Gamma)$ where $n$ is the number of
taps and $\Gamma$ is the strength of the tap. If one first obtain
a volume $V(n_1,\Gamma_1)$, and then repeat the experiment at a
different tap intensity and obtain $V(n_2,\Gamma_2)$, when we
return to tapping at $(n_1,\Gamma_1)$ one obtains a volume $V'$
which is $V'(n_1,\Gamma_1) = V(n_1,\Gamma_1)$. Moreover, in
simulations of slowly sheared granular systems the ergodic
hypothesis was shown to work \cite{nature} as we will discuss in
Section \ref{simulations}.

It is often noted in the literature that although the simple
concept of summing over all jammed states which occupy a volume
$V$ works, there is no first principle derivation of the
probability distribution of the Edwards ensemble as it is provided
by Liouville's theorem for equilibrium statistical mechanics of
liquids and gases. In granular thermodynamics there is no
justification for the use of the $\cal W$ function to describe the
system as Liouville's theorem justifies the use of the energy in
the microcanonical ensemble. In Section \ref{granularboltzmann} we
will provide an intuitive proof for the use of $\cal W$  in
granular thermodynamics by the analogous proof of the Boltzmann
equation.

The comment was also made that there is no proof that the entropy
Eq. (\ref{entropy}) is a  rigorous basis for granular statistical
mechanics. In the next section we will develop a Boltzmann
equation for jammed systems and show that this analysis can be
used to produce a second law of thermodynamics, $\delta S\ge 0$
for granular matter, and  the equality only comes with Eq.
(\ref{entropy}) being achieved.

Although everyone believes that the second law of thermodynamics
is universally true in thermal systems, the only accessible proof
comes in the Boltzmann equation, as the ergodic theory is a
difficult branch of mathematics which will not be covered in the
present discussion. By investigating the assumptions and key
points which led to the derivation of the Boltzmann equation in
thermal systems, it is possible to draw analogies for an
equivalent derivation in jammed systems.

It should be noted that there is an extensive literature on
granular gases \cite{savage,jenkins}, which are observed when
particles are fluidised by vigorous shaking, thus inducing
continuous particle collisions. There is  a powerful literature on
this topic, but it is not applicable to the problem of jamming.



\section{The Classical Boltzmann Equation}

Entropy in thermal systems satisfies the second law,

\begin{equation}
\label{secondlaw}\frac{\partial S}{\partial t} \ge 0,
\end{equation}
which states that there is a maximum entropy state which,
according to the evolution in Eq. (\ref{secondlaw}), any system
evolves toward, and reaches at equilibrium. A `semi'-rigorous
proof of the Second Law was provided by Boltzmann (the well-known
`H-theorem'), by making use of the `classical Boltzmann equation',
as it is now known.

\begin{figure}[tbp]
         \begin{center}
          \epsfig{file=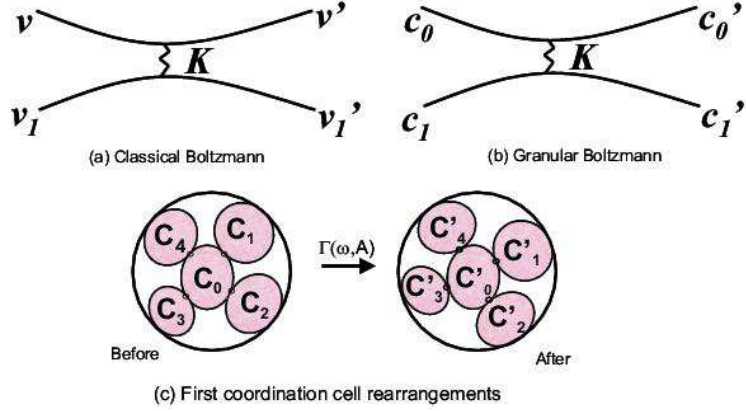,width=10 cm}
         \end{center}
\caption{(a) Collision of two particles in a dilute gas.
 (b) ``Collision of two configurations'' given in terms of
two contact points in a jammed material. (c) Rearrangements inside
a pocket of mobile grains under the first coordination shell
approximation for grain $\alpha=0$.\label{collision}}
\end{figure}

In order to derive this equation, Boltzmann made a number of
plausible assumptions concerning the interactions of particles,
without proving them rigorously. The most important of these
assumptions were:

\begin{itemize}
    \item The collision processes are dominated by two-body
    collisions (Fig. \ref{collision}a).
This is a plausible  assumption   for a dilute gas, since the
system
    is of very low density, and the
    probability of there being three or more particles colliding
    is infinitesimal.
    \item Collision processes are uncorrelated, i.e. all memory of
the collision is lost on completion and is not remembered in
subsequent collisions: the famous Stosszahlansatz. This is also
valid only for dilute gases, but the proof is more subtle.
\end{itemize}

Thus, Boltzmann proves Eq. (\ref{secondlaw}) for a dilute gas
only, but this is a readily available situation. The remaining
assumptions have to do with the kinematics of particle collisions,
i.e. conservation of kinetic energy, conservation of momentum, and
certain symmetry of the particle scattering cross-sections.

Let $f(v,r)$ denote the probability of a particle having a
velocity $v$ at position $r$. This probability changes in time by
virtue of the collisions. The two particle collision is visualised
in Fig. \ref{collision}a where $v$ and $v_1$ are the velocities of
the particles before the collision and $v'$ and $v'_1$ after the
collision.

On time scales larger than the collision time, momentum and
kinetic energy conservation apply:

\begin{equation}
m v + m v_1 = m v' + m v_1', \\\\\\\\\\\\\\\\\\\
\frac{1}{2} m v^2 +  \frac{1}{2} m v_1^2 = \frac{1}{2} m v'^2 +
\frac{1}{2} m {v'}_1^2. \label{conservation}
\end{equation}

Then, the distribution $f(v,r)$ evolves with time according to

\begin{equation}\label{fvr}
\frac{\partial f}{\partial t}
   + v \frac{\partial f}{\partial r} +
   \int {\cal K}(v, v' ; v_1, v_1')~~ \Big (
f(v) f(v_1) - f(v') f(v_1') \Big ) ~~ d^3v_1 d^3 v'd^3v_1' = 0.
\end{equation}

The kernel $\cal K$ is positive definite and
contains $\delta$-functions to satisfy the conditions
(\ref{conservation}), the flux of particles into the collision and
the differential scattering cross-section. We consider the case of
homogeneous systems, i.e. $f=f(v)$, and define
\begin{equation}
S = - k_B \int f \ln f.
\end{equation}
Defining $x = f f_1 / f' f'_1$ we obtain
\begin{equation}
\frac{\partial S}{\partial t} = \int {\cal  K}~  \ln x ~(1-x)
d^3 v_1 ~ d^3 v'~ d^3v_1',\\
(1-x) \ln x \ge 0, \\
{\cal K} \ge 0.
\end{equation}
Hence $ \partial S/ \partial t\ge0$ (see standard text books on
statistical mechanics).

It is also straightforward to establish the equilibrium
distribution where  $ \partial S/ \partial t = 0$ since it occurs
when the kernel term vanishes, i.e. when the condition of detailed
balance is achieved, $x=1$:
\begin{equation}\label{balance}
f(v) f(v_1) = f(v') f(v_1 ').
\end{equation}
The solution of Eq. (\ref{balance}) subjected to the condition of
kinetic energy conservation is given by the Boltzmann distribution
\begin{equation}
\label{boltzmann} {\displaystyle f(v) = \left (\frac{k_B T}{m\pi}
\right )^{3/2} ~   e^{-\frac{1}{2} \beta m v^2} },
\end{equation}
where $\beta = 1/k_B T$.
%
Equation (\ref{boltzmann}) is a reduced distribution and valid
only for a dilute gas. The Gibbs distribution represents the full
distribution and is obtained by replacing the kinetic energy in
(\ref{boltzmann}) by the total energy of the state to obtain:

\begin{equation}
\label{gibbs} P(E) \propto e^{-\beta E}.
\end{equation}

The question is whether a similar form can be obtained in a
granular system in which we expect

\begin{equation}
\label{canonical}
 P({\cal W}) \propto  e^{-{\cal W}/ \lambda X},
\end{equation}
where $X$ is the compactivity in analogy with
$T=\partial{E}/\partial{S}$. Such an analysis is shown in the next
section in an approximate manner.

\section{`Boltzmann Approach' to Granular Matter}
\label{granularboltzmann}

The analogous approach to granular materials consists in the
following: the creation of an ergodic grain pile suitable for a
statistical mechanics approach via a method for the exploration of
the available configurations analogous to Brownian motion, the
definition of the discrete elements tiling the granular system via
the volume function $\cal W$ (the sum of which provides the
analogous `Hamiltonian' to the energy in thermal systems), and an
equivalent argument for the energy conservation expressed in terms
of the system volume necessary for the construction of the
Boltzmann equation.

We have already established the necessity of preparing a granular
system adequate for real statistical mechanics so as to emulate
ergodic conditions. The grain motion must be well-controlled, as
the configurations available to the system will be dependent upon
the amount of energy/power put into the system. This pretreatment
is analogous to the averaging which takes place inherently in a
thermal system and is governed by temperature.

As explained, the granular system explores the configurational
landscape by the external tapping introduced by the
experimentalist. The tapping is characterised by a frequency and
an amplitude ($\omega, \Gamma$) which cause changes in the contact
network, according to the strength of the tap. The magnitude of
the forces between particles in mechanical equilibrium and their
confinement determine whether each particle will move or not. The
criterion of whether a particular grain in the pile will move in
response to the perturbation will be the Mohr-Coulomb condition of
a threshold force, above which sliding of contacts can occur and
below which there can be no changes. The determination of this
threshold involves many parameters, but it suffices to say that a
rearrangement will occur between those grains in the pile whose
configuration and neighbours produce a force which is overcome by
the external disturbance.

\begin{figure}[tbp]
        \centering
       \epsfig{file=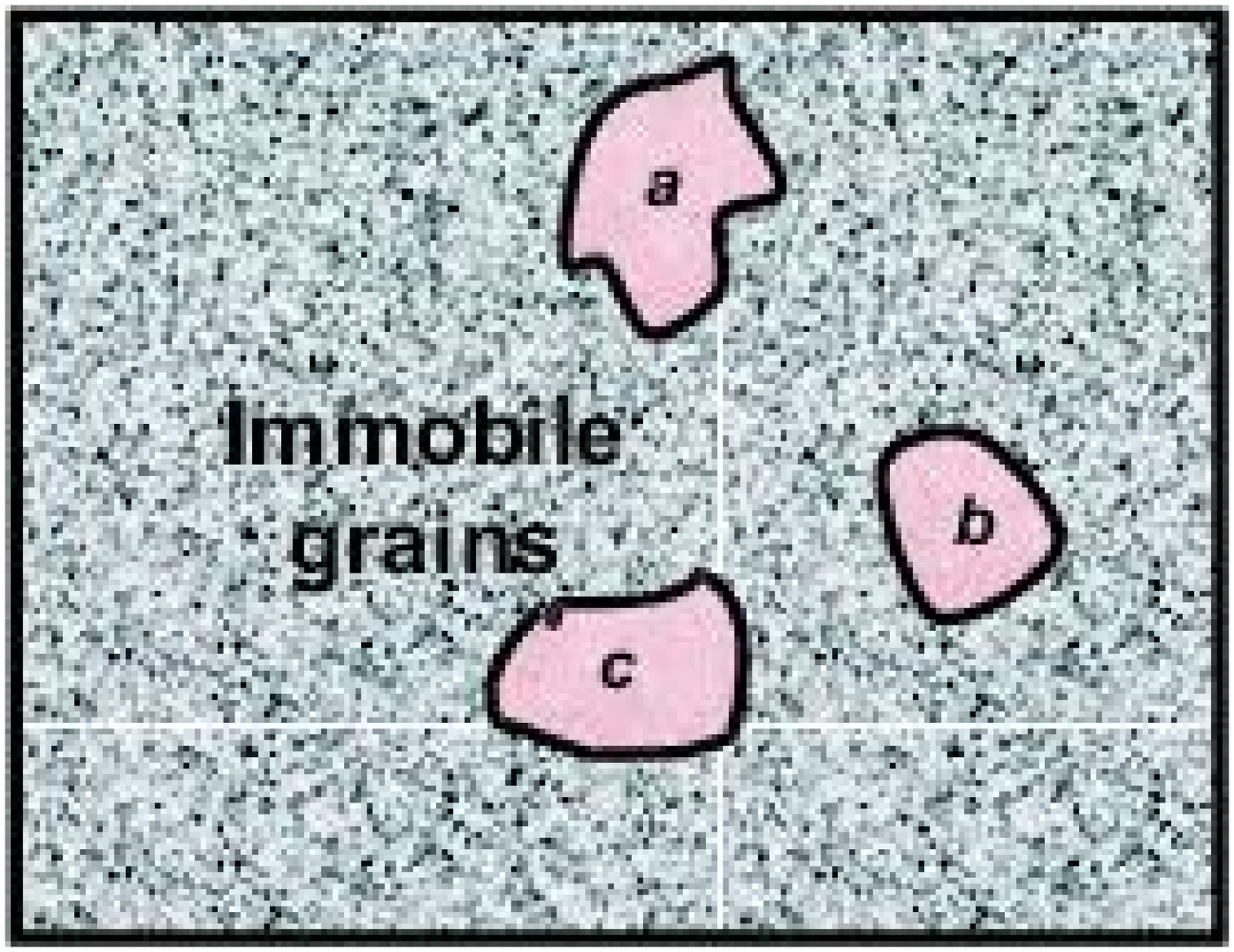,width=6 cm}
        \epsfig{file=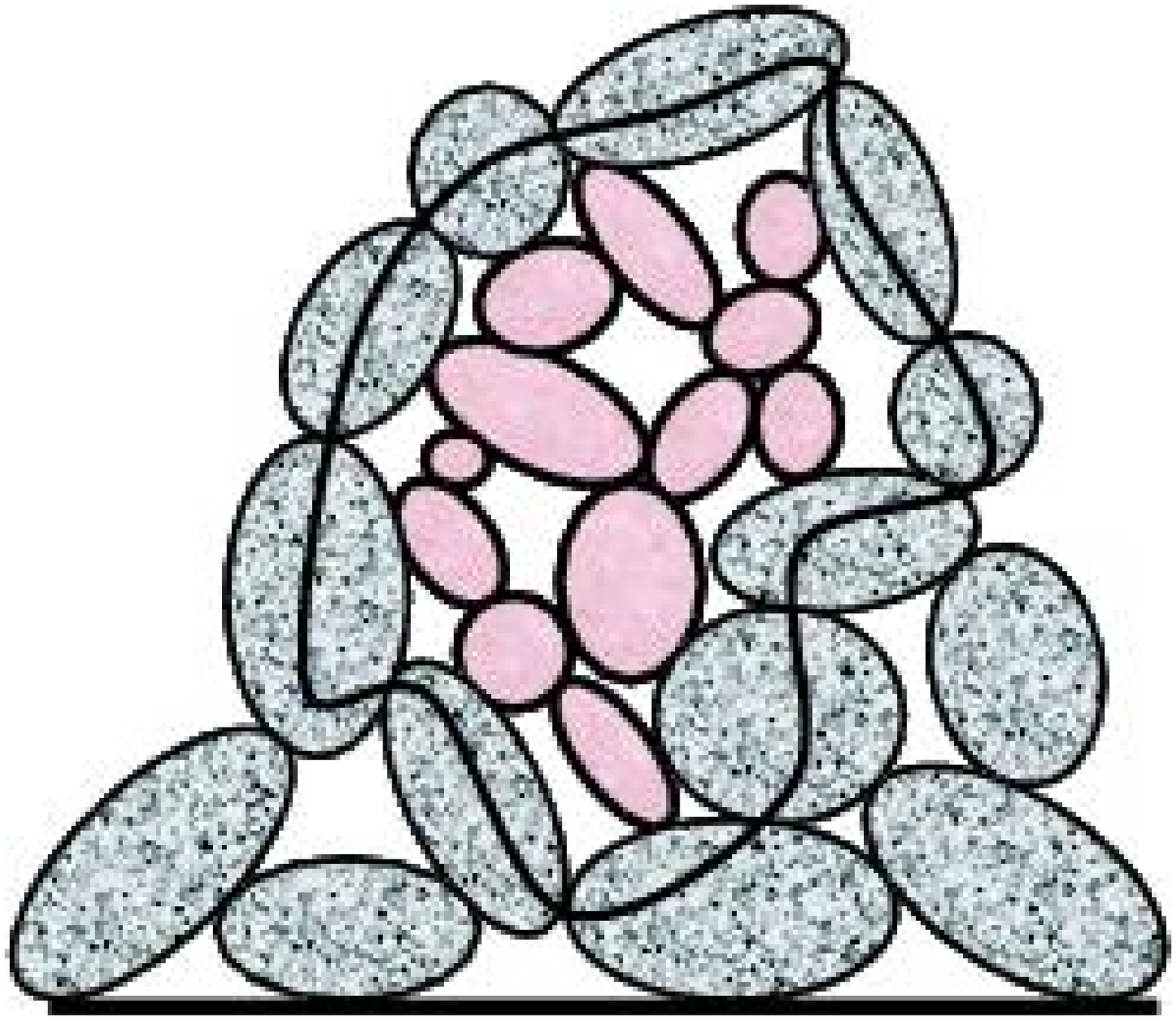,width=6 cm}
\caption{(a) Regions of mobile grains $a, b, c$ in a matrix of
immobile grains below the Coulomb threshold. (b) Detail of pocket
of mobile grains $a$ surrounded by immobile grains which are
shaded.} \label{pockets}
\end{figure}

The concept of a threshold force necessary to move the particles
implies that there are regions in the sample in which the contact
network changes and those which are unperturbed, shown in Fig.
\ref{pockets}.
Of course, since this is a description of a collective motion
behaviour, the region which can move may expand or contract, but
the picture at any moment in time will contain pockets of motion
encircled by a static matrix. Each of these pockets has a
perimeter, defined by the immobile grains. It is then possible to
consider the configuration before and after the disturbance inside
this well-defined geometry.

The present derivation assumes the existence of these regions. It
is equivalent to the assumption of a dilute gas in the classical
Boltzmann equation, although the latter is readily achieved
experimentally.

The energy input must be on the level of noise, such that the
grains largely remain in contact with one another, but are able to
explore the energy landscape over a long period of time. In the
case of external vibrations, the appropriate frequency and
amplitude can be determined experimentally for different grain
types, by investigating the motion of the individual grains or by
monitoring the changes in the overall volume fraction over time.
It is important that the amplitude does not exceed the
gravitational force, or else the grains are free to fly up in the
air, re-introducing the problem of initial creation just as they
would if they were simply poured into another container.

Within a region $a$ we have a volume $\sum_{\alpha \in a} {\cal
W}^\alpha$ and after the disturbance a volume which is now
$\sum_{\alpha \in a} {\cal W}'^\alpha$ as seen in Fig.
\ref{collision}b and \ref{collision}c.
 In Section \ref{W} we have discussed how to
define the volume function ${\cal W}^\alpha$ as a function of the
contact network. Here the simplest ``one grain'' approximation is
used as the ``Hamiltonian'' of the volume as defined by Eq.
(\ref{Wfunc}).
In reality it is much more complicated, and although there is only
one label $\alpha$ on the contribution of grain $\alpha$ to the
volume, the characteristics of its neighbours may also appear.

Instead of energy being conserved, it is the total volume which is
conserved while the internal rearrangements take place within the
pockets described above. Hence

\begin{equation}
\sum_{\alpha \in a}{{\cal W}^{\alpha}}=\sum_{\alpha \in a}{{\cal
W}'^{\alpha}}. \label{volumeconserved}
\end{equation}

We now construct a Boltzmann equation. Suppose $z$ particles are
in contact with grain $\alpha=0$, as seen in Fig.
\ref{collision}c. For rough particles $z=4$ while for smooth $z=6$
at the isostatic limit (see Section \ref{simulations}).
 The probability distribution will be of
the packing configurations which are represented by the tensor ${
\cal C}^\alpha$, Eq. (\ref{C}), for each grain, where $\alpha$
ranges from 0 to 4 in this case.
So the analogy of $f(v)$ for the Boltzmann gas equation becomes
$f({\cal C}^0)$ for the granular system and represents the
probability that the external disturbance causes a particular
motion of the grain. We therefore wish to derive an equation

\begin{equation}
\frac{\partial f({\cal C}^0)}{\partial t}
 +    \int {\cal K}({\cal C}^\alpha,{\cal C}'^\alpha)~~ \Big (
f_0 f_1 f_2 f_3 f_4 -  f'_0 f'_1 f'_2 f'_3 f'_4 \Big ) ~~d{\cal
C}'^0\prod_{\alpha\neq 0} d{\cal C}^\alpha  d{\cal C}'^\alpha = 0.
\end{equation}

The term $\cal K$ contains the condition that the volume is
conserved (\ref{volumeconserved}), i.e. it must contain
$\delta(\sum {\cal W}^\alpha - \sum {\cal W}'^\alpha)$.
The cross-section is now the compatibility of the changes in the
contacts, i.e. ${\cal C}^\alpha$ must be replaced in a
rearrangement by ${\cal C}'^\alpha$, Fig. \ref{collision}b (unless
these grains part and make new contacts in which case a more
complex analysis is called for). We therefore argue that the
simplest $\cal K$ will depend on the external disturbance $\Gamma,
\omega$ and on ${\cal C}^\alpha$ and ${\cal C}'^\alpha$, i.e.


\begin{equation}
\label{granularB} \frac{\partial f({\cal C}^0)}{\partial t} +
\int \delta \Big(\sum_\alpha {\cal
  W}^\alpha - \sum_\alpha {\cal W}'^\alpha\Big) ~
{\cal J}({\cal C}^\alpha,{\cal C}'^\alpha)~~ \Big (
 \prod_{\alpha=0}^{z} f_\alpha -  \prod_{\alpha=0}^{z} f'_\alpha
\Big )~ d{\cal C}'^0 \prod_{\alpha\neq 0} d{\cal C}^\alpha  d{\cal
C}'^\alpha
= 0.
\end{equation}
where $\cal J$ is the cross-section and it is positive definite.

The Boltzmann argument now follows. As before
\begin{equation}
S = -\lambda \int f \ln f,  \\ x = \frac{f_0 f_1 f_2 f_3 f_4}
{f_0' f_1' f_2' f_3' f_4'},
\end{equation}
and
\begin{equation}
\frac{\partial S}{\partial t} \ge 0,
\end{equation}
the equality sign being achieved when $x=1$ and
\begin{equation}
f_\alpha = \frac{e^{-{\cal W}^\alpha/\lambda X} }{Z},
\label{granu}
\end{equation}
with the partition function
\begin{equation}
Z = \sum_\alpha e^{-{\cal W}^\alpha/\lambda X} ~~~{\Theta_\alpha},
\end{equation}
and the analogue to the free energy being $Y = - X \ln Z$, and
$X=\partial V / \partial S$.

The detailed description of the kernel $\cal K $ has not been
derived as yet due to its complexity. Just as Boltzmann's proof
does not depend on the differential scattering cross section, only
on the conservation of energy, in the granular problem we consider
the steady state excitation externally which conserves volume,
leading to the granular distribution function, Eq. (\ref{granu}).

It is interesting to note that there is a vast and successful
literature of equilibrium statistical mechanics based on
$\exp(-{\cal H}/k_BT)$, but a meagre literature on dynamics based
on attempts to generalise the Boltzmann equation or, indeed, even
to solve the Boltzmann equation in situations remote from
equilibrium where it is still completely valid. It means that any
advancement in understanding how it applies to analogous
situations is a step forward.

\chapter{Jamming with the Confocal}
\label{experiments}

\section{From Micromechanics to Thermodynamics}

The first step in realising the idea of a general jamming theory
is to understand in detail the characteristics of jammed
configurations in particulate systems. Thus, the main aim of this
section is to design an experiment to provide a microscopic
foundation for the statistical mechanics of jammed systems. The
understanding of the micromechanics on the scale of the particle,
together with the respective statistical measures, pave the path
towards an experimental proof of the existence of such an
underlying thermodynamics.

The problem with the characterisation of the jammed state in terms
of its microstructure is that the condition of jamming implies an
optically impenetrable particulate packing. The fact that we
cannot take a look inside the bulk to infer the structural
features has confined all but one three-dimensional study of
packings to numerical simulations and the walls of an assembly. In
the old days Mason, a graduate student of Bernal, took on the
laborious task of shaking glass balls in a sack and `freezing' the
resulting configuration by pouring wax over the whole system. He
would then carefully take the packing apart, ball by ball, noting
the positions of contacts (ring marks left by the wax) for each
particle \cite{bernal}.
The statistical analysis of his hard-earned data led to the
reconstruction of the contact network in real space, a measurement
of the radial distribution function, $g(r)$, and also the number
of contacts of each particle satisfying mechanical equilibrium
which gives $z \approx 6.4$ for close contacts. This has been the
only reference point for simulators and theoreticians to compare
their results with those from the real world and it therefore
deserves a particular mention.

In search of an alternative method of experimentation, more in
line with the automated nature of our times, we developed a model
system suitable for optical observation. Moreover, our aim was to
investigate the jammed state in a different jammed medium, to
probe the universality of the configurational features. Finally,
one needs to solve the system geometry as well as the stresses
propagating through it in order to come up with a general theory.
To probe the stress propagation through the medium, rather than
its configuration alone, the particles must have well-defined
elastic properties. The system which could satisfy all our
requirements was found in a packing of {\it emulsion droplets}
\cite{faraday,behm}.

\section{Model system}

Emulsions are a class of material which is both industrially
important and exhibits very interesting physics \cite{princen}.
They belong to a wider material class of colloids in that they
consist of two immiscible phases one of which is dispersed into
the other, the continuous phase.
Both of the phases are liquids and their interface is stabilised
by the presence of surface-active species.

Emulsions are amenable to our study of jamming due to the
following properties:

\begin{description}

    \item[Transparency.] An alternative way of `seeing through' the
packing is to refractive index match the phases in the system -
i.e. the particles and the continuous medium filling the voids.
Since an emulsion is made up of two immiscible liquids, it is
possible to raise the refractive index of the aqueous phase to
match that of the dispersion of oil droplets. However,
transparency is not the only requirement, since the particles are
then dyed to allow for their optical detection.

    \item[Alternative Medium.] The emulsion is made up of smooth,
    stable droplets in the $1-10\mu$m size range, as compared to
    rigid, rough particles in the above described granular
    system. Both systems are athermal, but the length scale and the
    properties of the constituent particles of the system are very
    different.

    \item[Elasticity.] Emulsion droplets are deformable, stabilised by
an elastic surfactant film, which allows for the measurement of
the interdroplet forces from the amount of film deformation upon
contact. Moreover, the elasticity facilitates the measurement of
the dependence of the contact force network on the external
pressure applied to the system.

\end{description}

Our model system consists of a dense packing of emulsion oil
droplets, with a sufficiently elastic surfactant stabilising layer
to mimic solid particle behaviour, suspended in a continuous phase
fluid. The refractive index matching of the two phases, necessary
for 3D imaging, is not a trivial task since it involves
unfavourable additions to the water phase, disturbing surfactant
activity. The successful emulsion system, stable to coalescence
and Ostwald ripening, consisted of Silicone oil droplets
($\eta=10$cS) in a refractive index matching solution of water
($w_t=51\%$) and glycerol ($w_t=49\%$), stabilised by 20$mM$
sodium dodecylsulphate (SDS) upon emulsification and later diluted
to below the critical micellar concentration (CMC$=13mM$) to
ensure a repulsive interdroplet potential.
 The droplet phase is
fluorescently dyed using Nile Red, prior to emulsification. The
control of the particle size distribution, prior to imaging, is
achieved by applying very high shear rates to the sample, inducing
droplet break-up down to a radius mean size of $3.4\mu m$. This
system is a modification of the emulsion reported by Mason {\it et
al.} \cite{mason} to produce a transparent sample suitable for
confocal microscopy.

\subsection{Characterisation of a jammed state}

Having prepared a stable, transparent emulsion we use confocal
microscopy for the imaging of the droplet packings at varying
external pressures, i.e. volume fractions. The key feature of this
optical microscopy technique is that only light from the focal
plane is detected. Thus 3D images of translucent samples can be
acquired by moving the sample through the focal plane of the
objective and acquiring a sequence of 2D images.

\begin{figure}
\centerline {\hbox{
\includegraphics*[width=.8\textwidth]{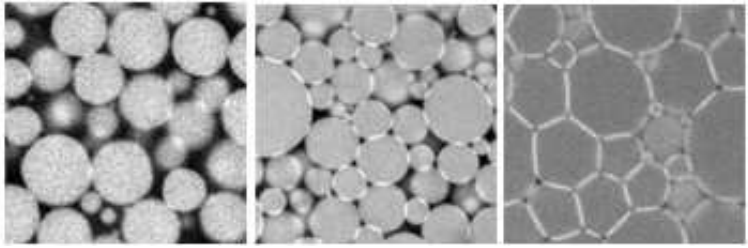}
} }
\caption{2D slices of emulsions under varying compression rates:
(a) 1g, (b) 6000g, and (c) 8000g.} \label{slices}
\end{figure}

Since the emulsion components have different densities, the
droplets cream under gravity to form a random close packed
structure. In addition, the absence of friction ensures that the
system has no memory effects and reaches a true jammed state
before measurement. If the particles are subjected to
ultracentrifugation, configurations of a higher density are
achieved as the osmotic pressure is increased. The random close
packing fraction reached under gravity depends on the
polydispersity of the emulsion, or in other words, the efficiency
of the packing. The sequence of images in Fig. \ref{slices} shows
2D slices from the middle of the sample volume after: (a) creaming
under gravity, (b) centrifugation at 6000g for 20 minutes and (c)
centrifugation at 8000g for 20 minutes. The samples were left to
equilibrate for several hours prior to measurements being taken.


The volume fraction for our polydisperse system shown in Fig.
\ref{slices}b is $\phi= 0.86$, determined by image analysis. This
high volume fraction obtained at a relatively small osmotic
pressure of 125 Pa is achieved due to the polydispersity of the
sample.






\begin{figure}[tbp]
         \begin{center}
          \epsfig{file=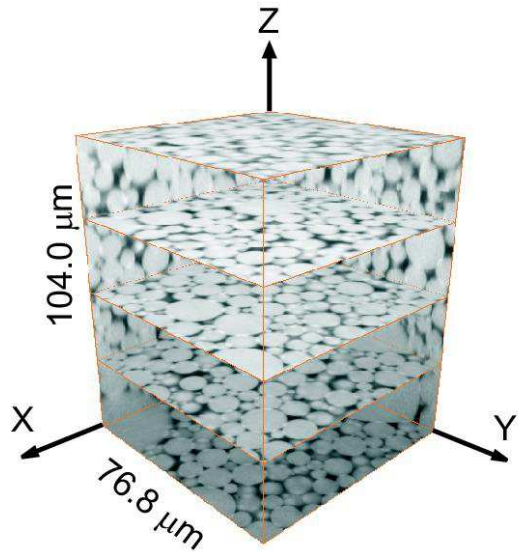,width=7 cm}
         \end{center}
\caption{Confocal image of the densely packed emulsion system.
\label{box}}
\end{figure}

\begin{figure}
\centerline {\hbox{
\includegraphics*[width=.72\textwidth]{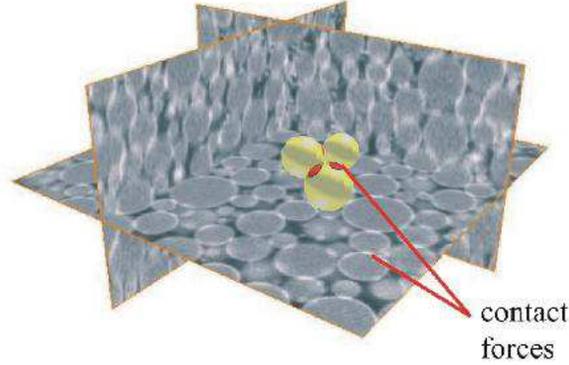}
} }
\caption{ A novel fluorescence mechanism using confocal microscopy
is providing new insight into the microstructure and the mechanics
of jammed matter. The suitable model system is an emulsion
comprised of oil droplets of approximately 3.4 $\mu$m dispersed in
a refractive index matching solution. The system forms a random
close packed structure by creaming or centrifugation giving rise
to a force network. Based on the bright contact areas, the
interdroplet forces can be extracted. The resulting micromechanics
is being used to develop statistical theories of jammed materials.
} \label{nature-fig}
\end{figure}

The 3D reconstruction of the 2D slices is shown in Fig. \ref{box}.
We have developed a sophisticated image analysis algorithm which
uses Fourier Filtering to determine the particle centres and radii
with subvoxel accuracy of all the droplets in the sample
\cite{faraday}. This data was previously unavailable from true 3D
experiments. Since the droplets are deformable and they exert
forces on one another upon contact, the area of droplet
deformation gives an approximation of the force. The deformed
areas appear brighter than the rest of the image due to an
enhanced fluorescence at the contact \cite{faraday}, allowing for
an independent measurement of the forces between droplets as can
clearly be seen in Fig. \ref{nature-fig}.

A system of random close packed particles is fully described by
the geometry of the system configuration and the distribution of
forces and stresses in the particulate medium. This means that if
$P$ is the probability distribution of configurations and of
interparticle forces, it consists of two independent components,

\begin{equation}
P = P_f(\mbox{forces}) P_c(\mbox{configurations}),
\end{equation}
which give the full description of the particulate system.

The above statement has been presented in a theory context, but
must be supported by experiment. In the next two sections we
present experimental results that test the basic granular theory
and some of the assumptions within it by separately measuring the
distribution of forces \cite{behm} and the distribution of
configurations \cite{capri}.

\subsection{Force distributions in a jammed emulsion}
\label{forcedistribution}

The micromechanics of jammed systems has been extensively studied
in terms of the probability distribution of forces, $P(f)$.
However, the experiments were previously confined to 2D granular
packs \cite{powders,baxter} or the measurement of the forces
exerted at the walls of a 3D granular assembly
\cite{chicago3,Mueth1998,Lovoll1999,Blair2001,Makse2000} thus
reducing the dimensionality of the problem. On the other hand,
numerical simulations \cite{Radjai1996,Thornton,Makse2000,ohern}
and statistical modelling \cite{Coppersmith95} have provided the
$P(f)$ for a variety of jammed systems, from structural glasses to
foams and compressible particles, in 3D. Our novel experimental
technique can be compared with all previous studies in search of a
common behaviour. Apart from being the first study of $P(f)$ in
the bulk, it is also the first study of jamming in an emulsion
system.

\begin{figure}
\centerline {\hbox{
\includegraphics*[width=.6\textwidth]{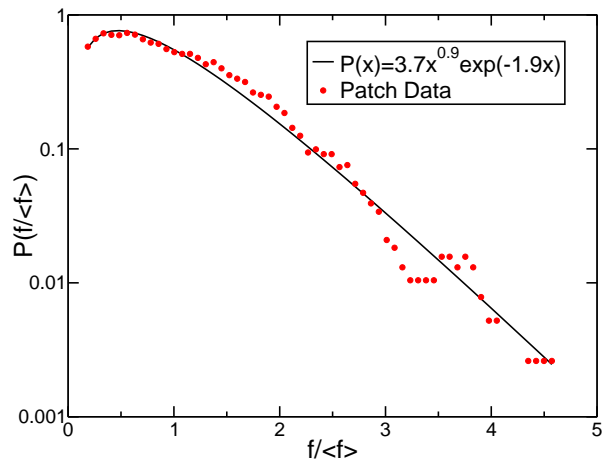}
}} \caption{Probability distribution of the contact forces for the
compressed emulsion system shown in Fig. \ref{slices}b. We also
show a fit to the theory developed in \protect\cite{behm}.}
\label{pf}
\end{figure}

The result of the confocal images analysis, shown in Fig.
\ref{pf}, shows the probability distribution of interdroplet
forces, $P(f)$, for the sample shown in Fig. \ref{slices}b. We use
a linear force model (see Section \ref{protocol}) to obtain the
interdroplet forces from the contact area data extracted from the
image analysis described above.

The distribution data shown are extracted from 1234 forces arising
from 450 droplets. The forces are calculated from the bright,
fluorescent patches that highlight the contact areas between the
droplets as seen in Fig. \ref{nature-fig}. The droplets are only
slightly deformed away from spherical at the low applied
pressures. This indicates that the system is jammed near the RCP.

The data shows an exponential distribution at large forces,
consistent with results of many previous experimental and
simulation data on granular matter, foams, and glasses. The
quantitative agreement between $P(f)$ of a variety of systems
suggests a unifying microstructural behaviour governed by the
jammed state. The salient feature of $P(f)$ in jammed systems is
the exponential decay above the mean contact force. The behaviour
in the low force regime indicates a small peak, although the power
law decay tending towards zero is not well pronounced. The best
fit to the data gives a functional form of the distribution (see
Fig. \ref{pf}):
\begin{equation}
P(f)\propto {f}^{0.9}e^{-1.9f/\bar{f}},
\end{equation}
with $\bar{f}$ is the mean force. This form is consistent with the
theoretical q-model \cite{Coppersmith95} and with the model
proposed in \cite{behm} which predicts a general distribution of
the form $P(f)\propto f^n e^{-(n+1)f/\bar{f}}$, where the power
law coefficient $n$ is determined by the packing geometry of the
system and the coordination number.


\begin{figure}
\centerline {\hbox{
\includegraphics*[width=.6\textwidth]{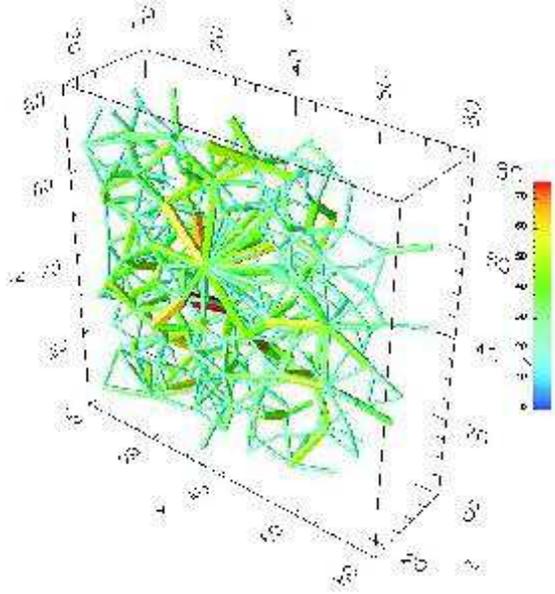}}}
\caption{ Force chains in emulsions: Plot of the interdroplet
forces inside the packing of droplets obtained in the experiments.
We plot only the forces larger than the average for better
visualisation. Each rod joining the centers of two droplets in
contact represents a force. The thickness and the colour of the
rod is proportional to the magnitude of the force.
 } \label{chains1}
\end{figure}

\begin{figure}
\centerline {\hbox{(a)
\hbox{\includegraphics*[width=.6\textwidth]{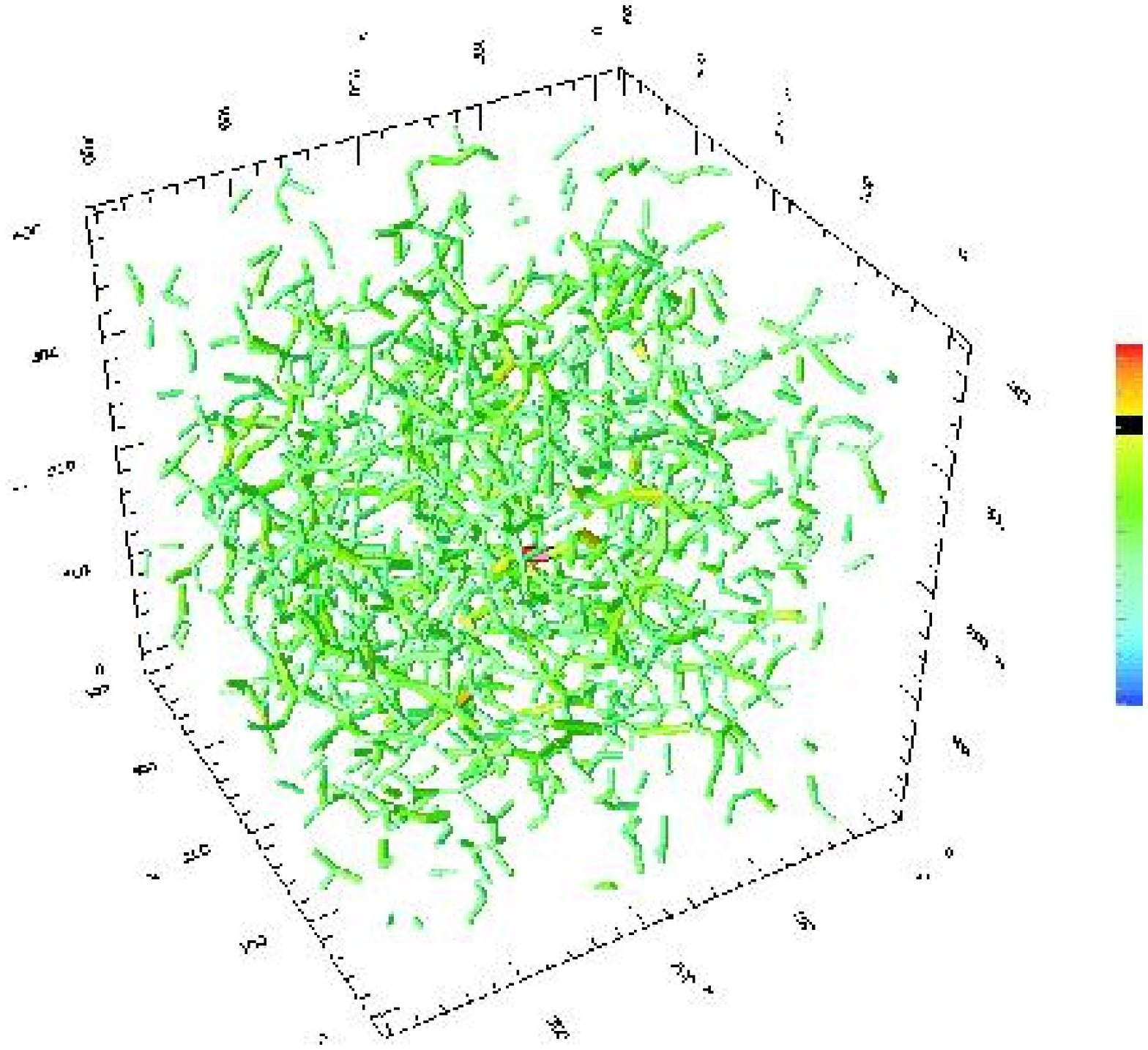} } } }
\centerline{ (b)
\hbox{\includegraphics*[width=.37\textwidth]{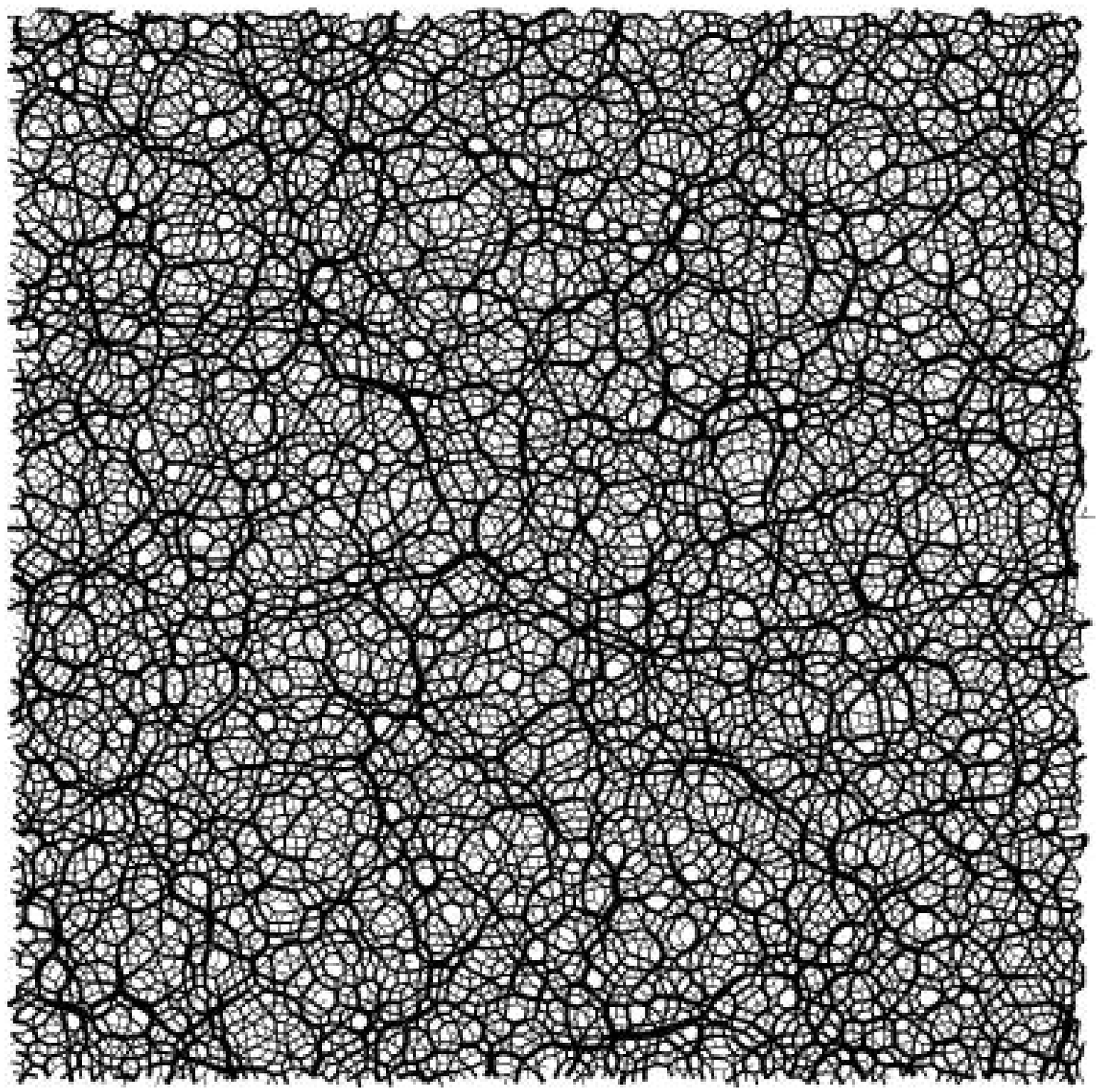} (c)
\includegraphics*[width=.6\textwidth]{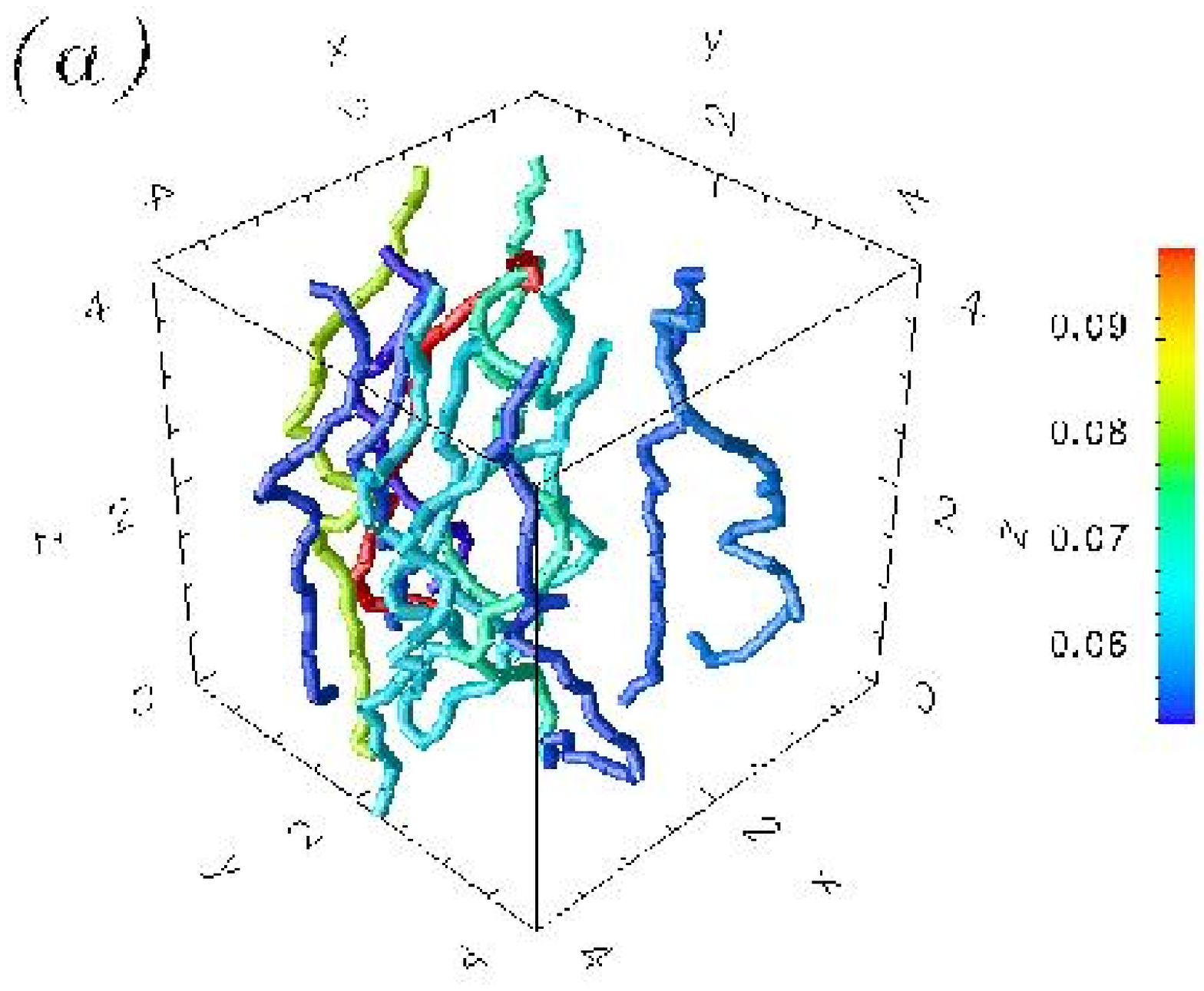} }} \caption{
Force chains in granular matter. (a)
 Frictionless isotropic granular system at $p=100$ KPa in 3D
from simulations. We plot only the forces larger than the average.
Force chains are
 tenuous and not well defined. (b)
Clearly visible force chains in a 2D frictional system from
simulations. (c) Frictional system under uniaxial compression from
simulations. Percolating force chains are seen in this case.}
\label{chains2}
\end{figure}

Our experimental data allows us to examine the spatial
distribution of the forces in the compressed emulsion, shown in
Fig. \ref{chains1}. In this sample volume, the forces appear to be
uniformly distributed in space and do not show evidence of
localisation of forces within the structure. Moreover, we find
that the average stress is independent of direction, indicating
isotropy. For comparison, we also show computer simulation results
for isotropic packings of Hertz-Mindlin spherical particles in
Fig. \ref{chains2} (see Section \ref{simulations}) where force
chains are not prominent either. On the other hand 2D packings
clearly show the existence of force chains under isotropic
pressure, indicating that their existence may be related to the
dimensionality of the problem. Furthermore, force chains can be
obtained in 3D by uniaxially compressing an isotropic packing, as
shown in Fig. \ref{chains2}c. It should be noted that in the case
of Fig. \ref{chains2}c an algorithm which {\it looks} for force
chains is applied by starting from a sphere at the top of the
system, and following the path of maximum contact force at every
grain. Only the paths which percolate are plotted, i.e., the
stress paths spanning the sample from the top to the bottom.

Interestingly, the salient feature of all the particulate packings
shown in Figs. \ref{chains1} and
 \ref{chains2},
irrespective of their spatial characteristics, is an exponential
distribution of forces. This indicates that force chains are not
necessary to obtain such a distribution. The rationalization of
this observation has been  exploited in the theory developed in
\cite{behm}
 which is based on the assumption of uncorrelated
force transmission through the packing.

The mean coordination number of the system, $Z$, is another
theoretically important parameter. It has been shown that the
isostatic limit is achieved (in 3D) for $Z = 4$ for frictional
systems and $Z = 6$ for smooth particles (see Section
\ref{isostatic}). This has not been tested in the real world until
present, except in the famous experiment of Bernal for smooth
particles who obtained a coordination number for spheres in
contact of 6.4 for metallic balls of 1/8' diameter. Even there,
all particles in contact were counted, whereas the theory predicts
the coordination number assuming only those particles which are
exerting a force. Our experiments at low confining pressures (up
to 200Pa) show that the mean coordination number is $Z\approx 6$
which can be interpreted as the isostatic limit for frictionless
spheres.


This completes the study of the jammed structures in terms of the
force distributions. There are more subtle ways in which the
static structure of the configuration can be investigated as a
statistical ensemble as we describe in the next section.

\subsection{Experimental measurement of $\cal W$ and $X$}
\label{measuringX}

Having investigated the probability distribution of forces within
the system, we now consider the distribution of the configurations
of the packing. In Section \ref{theory} we have justified the
application of statistical mechanics to jammed conditions,
provided there is a mechanism for changing the configurations by
tapping. The probability distribution of configurations is
governed by Eq. (\ref{canonical}).

\begin{figure}[tbp]
         \begin{center}
          \epsfig{file=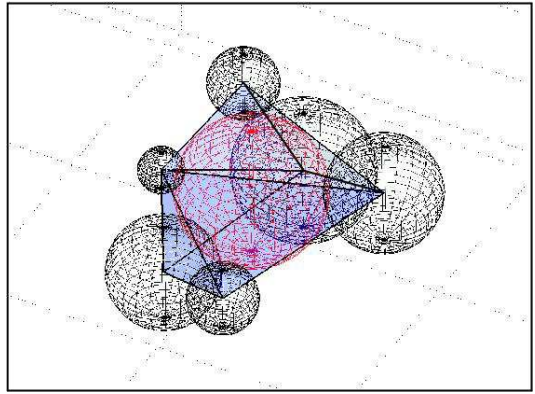,width=7 cm}
         \end{center}
\caption{An example of a volume $\cal W$ as the polyhedron
constructed from the 3D images. The centre grain (red) has 6
grains in contact (black), the centres of which are joined to form
the polyhedron. \label{Wpoly}}
\end{figure}

Using an extension to the same image analysis method used to
measure the distribution of forces, the 3D images of a densely
packed particulate model system also allow for the
characterisation of the volume function $\cal {W}$. This is
performed by the partitioning of the images into first
coordination shells of each particle, described in Section
\ref{W}. The polyhedron obtained by such a partitioning is shown
in Fig. \ref{Wpoly} and its volume is calculated from Eq.
(\ref{Wfunc}).



The ability to measure this function and therefore its
fluctuations in a given particle ensemble, enables the
calculations of the macroscopic variables. We calculate the
probability distribution of the volume per particle in the whole
image and find an exponential behaviour:

\begin{equation}
P({\cal W}) \propto e^{-{\cal W}/\lambda X},
\end{equation}

\begin{figure}[tbp]
         \begin{center}
          \epsfig{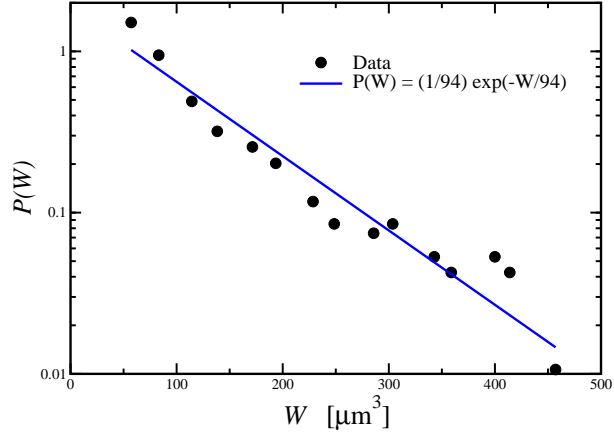}
         \end{center}
\caption{Probability distribution of $\cal W$ fitted with a single
exponential. The decay constant is the compactivity, $X = 94 \mu
m^3/\lambda$. \label{distW}}
\end{figure}

The exponential probability distribution of $\cal {W}$ enables the
measurement of the compactivity $X$ according to Eq.
(\ref{canonical}). The value obtained in this way is
$X=94\mu$m$^3/\lambda$, shown in Fig. \ref{distW}. The conversion
of this measurement of the compactivity into a measurement of the
analogue of temperature requires a new temperature scale for
granular matter. In other words, $\lambda$ (the analogue of the
Boltzmann constant in thermal systems) provides the link between
volume fluctuations (energy) and compactivity (temperature) and
needs to be determined for jammed matter.

We have shown that we can arrive at the thermodynamic system
properties from the knowledge of the microstructure. Many images,
i.e. configurations, can be treated in this way to test whether
system size influences the macroscopic observables. If the
particles are subjected to ultracentrifugation resulting in
configurations of a higher density, the influence of pressure on
the macroscopic variables can also be tested.

Such a characterisation of the governing macroscopic variables,
arising from the information of the microstructure, allows one to
predict the system's behaviour through an equation of state. This
is the first experimental study of such statistical concepts in
particulate matter and opens new possibilities for testing the
above described thermodynamic formulation. In principle, one can
apply low amplitude vibrations to the system and observe the
droplet configuration before and after the perturbation, thus
testing the ideas proposed in the Boltzmann derivation.

\chapter{Jamming in a Periodic Box}

\label{simulations}

The following section describes the potential for using computer
simulations in testing the thermodynamic foundations raised in the
previous sections. Rather than employing rough rigid grains for
which most of the theoretical concepts have so far been devised,
computer simulations are obliged to introduce some deformability
into the constituent particles to facilitate the measurement of
the particle interactions with respect to their positions. As a
result, the entropic considerations which have been explained only
in terms of the volume in the case of rigid grains in Section
\ref{theory} will now be generalised to situations in which there
is a finite energy of deformation in the system. The energy of the
system will parametrise the compaction curves at varying confining
pressures shown in Fig. \ref{comp}.

The entropy Eq. (\ref{entropy}) can then be redefined as a
function of both energy and volume,
\begin{equation}
S(E,V) = \lambda \ln \Sigma_{\mbox{\scriptsize jammed}}(E,V).
\end{equation}

The introduction of energy into the system implies a corresponding
compactivity,

\begin{equation}\label{compactivityE}
X_E^{-1} = \frac{{\partial S} }{{\partial E}},
\end{equation}
where the subscript $E$ denotes that the compactivity is now the
Lagrange multiplier controlling the energy of the jammed
configuration, not the volume. Notice that $X_E$ differs from the
temperature of an equilibrium system $T = \partial E/\partial S$
because the energy in Eq. (\ref{compactivityE}) is the energy of
the jammed configurations and not the thermal equilibrium energy.
The assumptions of ergodicity and equally probable microstates for
a given energy and volume are still valid here, just as they were
for rigid grains in the previous sections.

The canonical distribution in Eq. (\ref{canonical}) is generalised
to

\begin{equation}
\label{canonical2}
    P_\nu = \frac{e^{(-E_\nu/\lambda X_E - {\cal W}_\nu/\lambda X_V )} ~~ \Theta_\nu}
{Z} .
\end{equation}
The term $\Theta_\nu$ assures that we are considering only the
jammed configuration. Its significance will be discussed in
Section \ref{ergodicity}.

\section{Simulating Jamming}
\label{protocol}

 In Section \ref{granularjamming} we described the
need for a true jammed configuration before any statistical
measurement can be applied. While this process is achieved via
external perturbations in a laboratory experiment, the equivalent
procedure guaranteeing reproducible results using simulations,
requires a particular `equilibration' procedure. At each pressure,
the grains are pretreated in the following way in order to ensure
that all memory effects have been lost in the system.

We perform Molecular Dynamics (MD)  simulations of an assembly of
spherical grains in a periodically repeated cubic cell. The system
is composed of soft elasto-frictional spherical grains interacting
via Hertz-Mindlin contact forces, Coulomb friction and viscous
dissipative terms. Two model systems are investigated: granular
materials and compressed emulsions.

{\it Granular matter model.--} Particles are modelled as
viscoelastic spheres with different coefficients of friction.
Interparticle forces are computed using the principles of contact
mechanics \cite{johnson}. Full details are given in Refs.
\cite{Makse2000,mgjs,nature}. The normal force $F_n$ has the
typical 3/2 power law dependence on the overlap between two
spheres in contact (Hertz force), while the transverse force $F_t$
depends linearly on the shear displacement between the spheres, as
well as on the value of the normal displacement
 (Mindlin tangential elastic force).
As the shear displacement increases, the elastic tangential force
$F_t$ reaches its limiting value given by Amonton's law for no
adhesion, $F_t \leq \mu_f F_n$, which is a special case of
Coulomb's law. Viscous dissipative forces, proportional to the
relative normal and tangential velocities of the particles, are
also included to allow the system to equilibrate.

A granular system with tangential elastic forces is path-dependent
since the  work done in deforming the system depends upon the path
taken and not just the final state. On the other hand, a system of
spheres interacting only via normal forces is said to be
path-independent, and the work does not depend on the way the
strain is applied. It turns out that this is a good model for a
compressed emulsion system since they do not exhibit frictional
forces.

{\it Compressed emulsions model.--} A system of {\it frictionless}
viscoelastic spherical particles could be thought of as a model of
compressed emulsions \cite{lacasse,durian}, see also
\cite{faraday} for details. Even though they can be modelled in
this way, an important difference arises in the interdroplet
forces, which are not given in terms of the bulk elasticity, as
they are in the Hertz theory.
Instead, forces are given by the principles of interfacial
mechanics \cite{princen}. For small deformations with respect to
the droplet surface area, the energy of the applied stress is
presumed to be stored in the deformation of the surface. The
simplest approximation considers an energy of deformation which is
quadratic in the area of deformation \cite{princen}, analogous to
a harmonic oscillator potential which describes a spring
satisfying Hooke's law.

\begin{figure}
\centerline {\hbox{
\includegraphics*[width=.4\textwidth]{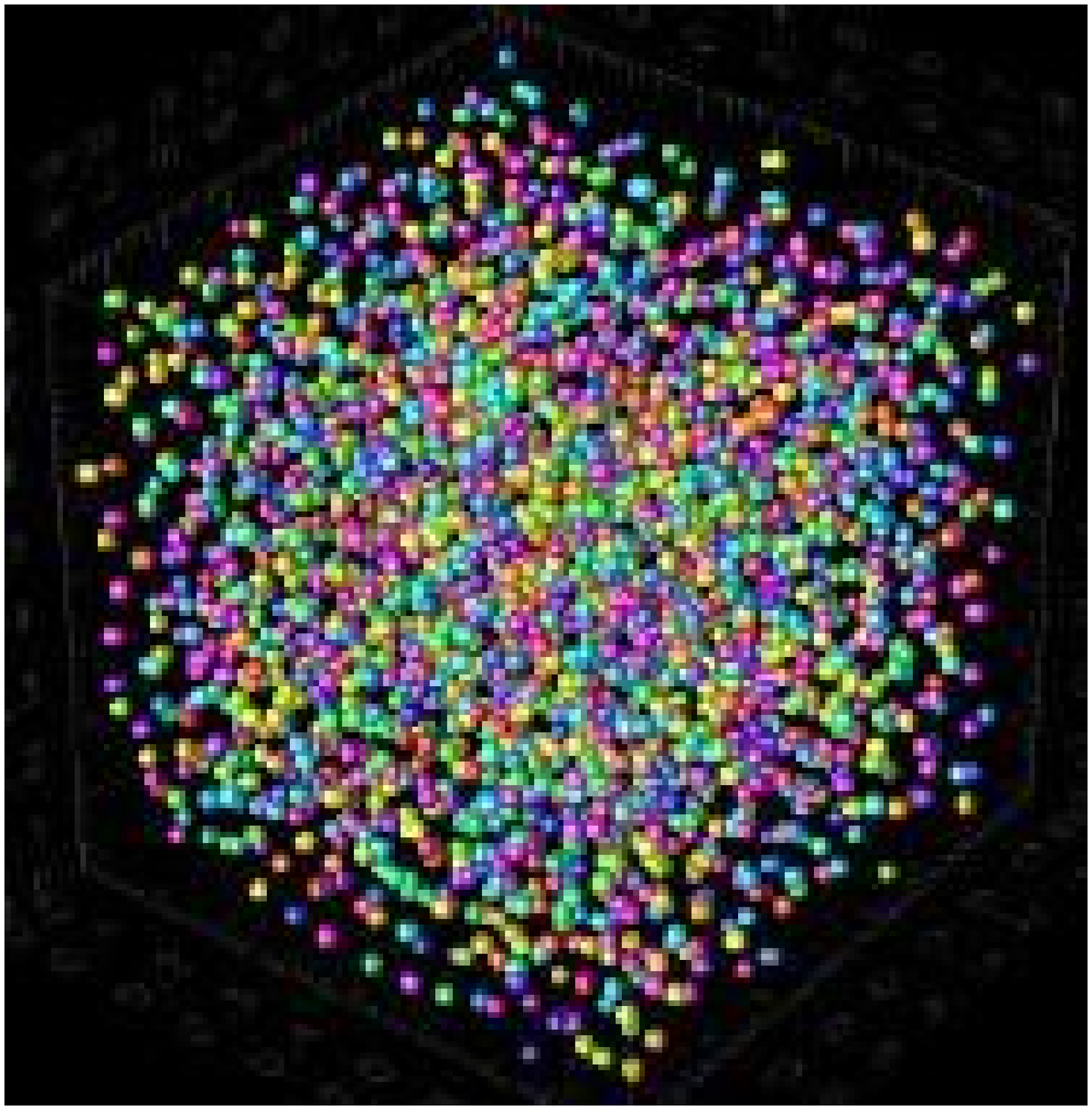}
 \includegraphics*[width=.4\textwidth]{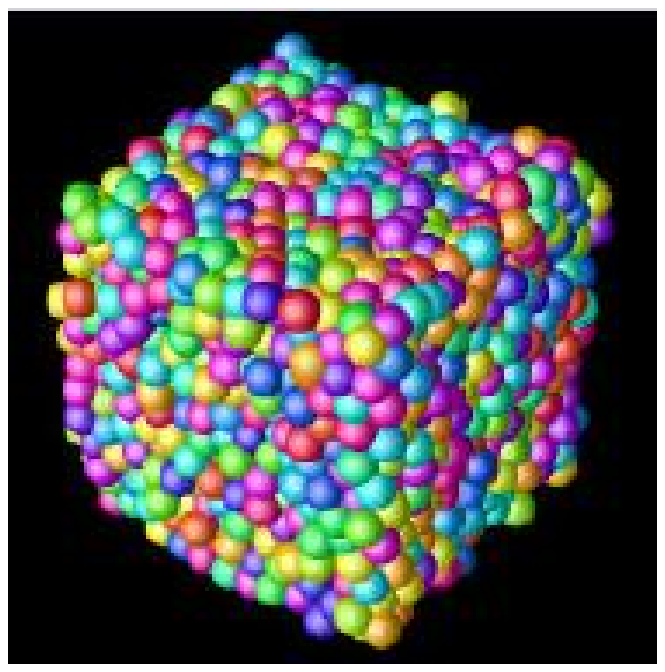} } }
\caption{Preparation protocol for granular materials and droplets.
We start with a system of noninteracting spheres (a) and then we
apply a compression protocol to reach the jammed state (b)}
\label{validity}
\end{figure}

There have been several more detailed numerical simulations
\cite{lacasse} to improve on this model and allow for
anharmonicity in the droplet response by also taking into
consideration the number of contacts by which the droplet is
confined. Typically these improved models lead to a force law for
small deformations of the form $F_n \propto A^b$ , where $A$ is
the area of deformation and $b$ is a coordination number dependent
exponent ranging from 1 (linear model) to 3/2 (Hertz model). For
simplicity and for a better comparison with the physics of
granular materials, in the following we will show results  only
considering the nonlinear 3/2 dependence of the normal force. A
more realistic nonlinear dependence is considered elsewhere in
\cite{part2}.

Thus we adjust the MD model of granular materials to describe the
system of compressed emulsions by only excluding the  transversal
forces (tangential elasticity and Coulomb friction). The
continuous liquid phase is modelled in its simplest form, as a
viscous drag force acting on every droplet, proportional to its
velocity.

{\it Preparation protocol.---} Our aim is to introduce a numerical
protocol designed to mimic the experimental procedure used to
achieve the jammed state, as explained in Section
\ref{jammedstate}. The simulations begin with a gas of
noninteracting grains distributed at random positions in a
periodically repeated cubic cell, depicted in Fig. \ref{validity},
showing snapshots of our typical simulations with 10,000 particles
of size $R=100 \mu$m. To avoid issues of path-dependency
introduced by the shear forces, the transverse force between the
grains is excluded from the calculation ($F_t = 0$). Because there
are no transverse forces, the grains slip without resistance and
the system reaches the high volume fractions found experimentally,
thus avoiding the irreversible branch of the compaction curve.
This procedure essentially mimics the path to jammed states for
the compressed emulsion system.

\begin{figure}
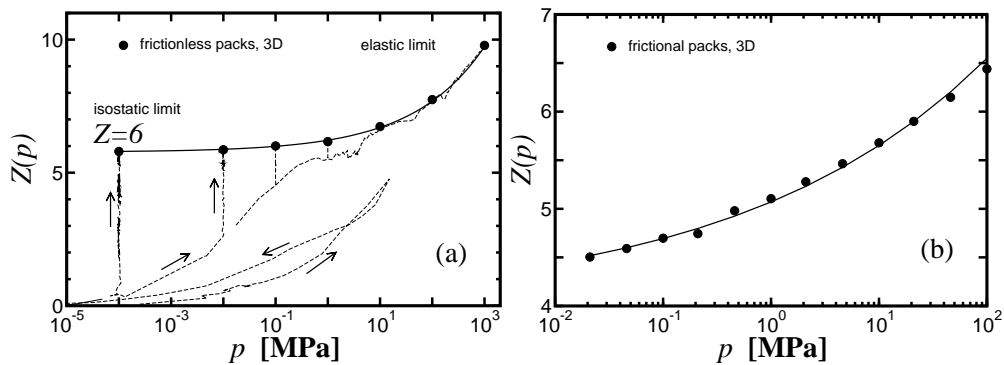

\centerline {\hbox{
\includegraphics*[width=.5\textwidth]{fig3a.eps}
 \includegraphics*[width=.5\textwidth]{fig3b.eps}} }
\caption{Coordination number versus pressure. (a) Frictionless
packs in 3D ($\mu=0$).  The system is isostatic  $Z_c=2D=6$ as
$p\to 0$. (b) Frictional packs in 3D (finite $\mu$).
} \label{coord}
\end{figure}

The protocol is then repeated for grains with friction. Initially,
a fast compression of the grains brings the system to the
irreversible branch of the compaction curve. It is then necessary
to apply a compression protocol in order to reach a target
pressure. This pressure is maintained with a ``servo'' mechanism
by the continuous application of an oscillatory strain until the
system reaches the jammed state \cite{cundall}. The servo
mechanism is analogous to the application of a small tapping
amplitude to reach the reversible branch of the compaction curve,
Fig. \ref{compaction}. In general, we find that by preparing the
system with frictional and elastic tangential forces, the system
reaches states of lower volume fractions. At the end of the
preparation protocol (depicted by the dashed lines in Fig.
\ref{coord}a) we obtain a set of jammed systems at different
stresses (for granular materials) or osmotic pressures (for
droplets) \cite{Makse2000}.

\subsection{Isostatic jamming}
\label{isostatic}

Consider a static packing of grains under an external force
$\vec{f}^{\mbox{\scriptsize ext}}$. The internal stresses obey the
Cauchy equations:
\begin{equation}
\partial_i \sigma_{ij} +f_j^{\mbox{\scriptsize ext}} =0.
\label{cauchy}
\end{equation}
Since there are 3 equations (in 3-D) for 6 independent stress
components (the stress tensor is symmetric), then the system is
indeterminate, and the Cauchy equations must be augmented by
additional constitutive equations. The conventional elastic
approach is then to consider the deformability of the packing
which is described by the strain field. Linear constitutive
relations are introduced to relate the strain to the stress via
the elastic constants of the material (Hooke's law). For an
isotropic elastic body only two elastic constants (for instance,
the shear modulus and the Poisson ratio) are sufficient to fully
describe the stress transmission in an elastic packing
\cite{landau-elastic}.

In the limit of infinitely rigid grains the strain field is
ill-defined and the validity of elasticity theory is open to
debate. In this case, it has been argued that it is possible to
solve the stress distribution,  based on Newton's equations alone
without resorting to the existence of the strain, only when the
system is at a particular {\it minimal coordination number}
\cite{edwards-grinev,bouchaud,ball}. The granular indeterminacy is
then solved by resorting to the configurational information alone.
Such approaches are intimately related to the thermodynamics of
jamming \cite{blumenfeld}.

The minimal coordination number can be understood in terms of
simple constraint arguments for a system of $N$ rigid spherical
grains in $D$ dimensions
\cite{alexander,isostatic,edwards-grinev}. In the case of
frictionless grains, $Z N/2$ normal forces have to be determined
with $D N$ equations of force balance. The critical coordination
number for which the equations of force balance are soluble is
fount to be $Z_c=2 D$. Similar arguments lead to a minimal
coordination of $Z_c = D+1$ for infinitely rough spherical grains,
i.e. grains with finite tangential forces, $F_t$,  but with an
infinitely large friction coefficient ($\mu \to \infty$).

Below $Z_c$ the system cannot be jammed and it exists only in
suspensions. Above $Z_c$ the system is underconstrained and
elasticity theory may give the correct approach to describe such a
packing of deformable grains. At the minimal coordination number
the system is in a state of marginal rigidity \cite{ball},
otherwise known as the isostatic limit \cite{alexander,isostatic}.

In order to test the existence of the isostatic limit we study the
coordination number dependence on pressure for the two cases:
frictionless grains and those with friction. The preparation
protocols explained above are performed to achieve different
target pressures and we obtain the average coordination number
$Z(p)$ of the jammed states as a function of the pressure, as
shown in Fig. \ref{coord}.

In the case of frictionless grains we find that the coordination
number of the pack approaches the minimal value $Z_c\approx 6$ as
$p \to 0$. At low pressures compared to the shear modulus of the
beads ($p\ll 26$ GPa) the system behaves most like a pack of rigid
balls, thus approaching the isostatic limit. The same preparation
protocol gives $Z_c\approx 4$ in 2D confirming the  $Z_c=2D$
relationship (results not shown here). The preparation protocol
for grains with infinite friction in 3D gives rise to different
packings with lower coordination numbers.
Our results suggest that the isostatic limit $Z_c=4$ is achieved
for infinitely rough grains as $p\to 0$ \cite{part2}. Grains with
{\it finite} friction also seem to tend to the minimal
coordination number $Z_c\approx 4$ as seen in Fig. \ref{coord}b,
but at much smaller pressures than those achieved in our
calculations. Therefore, it is difficult to draw conclusions as to
whether the isostatic limit requires even lower confining
pressures or not. In fact, recent computational studies have
suggested that the isostatic limit may not exist in packs with
finite friction \cite{silbert}. However, this study uses a
different compression protocol more akin to a fast quench, which
may leave the system trapped in the irreversible branch of the
compaction curve due to the system's inability to explore all the
available configurational space.

The approach to the marginal rigidity state $Z\to Z_c$ can be seen
as a jamming transition between a solid-like state with a finite
shear modulus and a liquid-like state with no resistance to shear,
observed in suspensions. In fact we find that the stress $\sigma$
and the shear modulus $G$ of the packing vanish according to a
power law as the system approaches a critical density, $\phi_c$,
corresponding to the jamming transition \cite{Makse2000}:


\begin{equation}
\sigma \sim (\phi - \phi_c)^\alpha, ~~~~~~~~ G \sim (\phi -
\phi_c)^\beta.
\end{equation}

The exponents $\alpha$, $\beta$ can easily be calculated in terms
of the microscopic law of interparticle interactions. For
instance, Hertz theory predicts the values of $\alpha = 3/2$ and
$\beta =1/2$, in agreement with with our simulation results. The
critical density $\phi_c$ depends on the interaction potential
between the grains.
A value $\phi_c=0.63\approx \phi_{\mbox{\scriptsize RCP}}$ is
achieved for frictionless grains and corresponds to the volume
fraction at RCP \cite{bernal}. On the other hand $\phi_c <
\phi_{\mbox{\scriptsize RCP}}$ are achieved for grains with
friction. These states correspond to RLPs. The jamming transition
can be thought of as a particular second-order phase transition
because the exponents are not universal; they depend on the
details of the microscopic interactions between the grains
\cite{ohern2}.

The coordination number also approaches the critical minimal value
as a power law. Empirically, we find (Fig. \ref{coord})
\begin{equation}
Z(p) = Z_c + \left(\frac{p}{ ~\mbox{6.45 MPa}}\right)^{1/3}.
\label{Z}
\end{equation}
with $Z_c=6$ and $Z_c=4$ for the infinitely smooth and infinitely
rough grains, respectively.

After having characterised the jammed state, the computational
study proceeds to develop the thermodynamics using the states
depicted in Fig. \ref{coord} as the starting point.

\section{Testing the Thermodynamics}
\label{testing}

If it were true that a thermodynamic framework could describe the
behavior of jammed systems, it stands to reason that the
compactivity $X_E$ of the granular pack can be measured from a
dynamical experiment involving the exploration of the energy
landscape. We examine the validity of this statement with computer
simulations in the following discussion.

Following the equilibration procedure, the exploration of the
energy landscape equivalently needs a driving mechanism such that
all appropriate configurations are sampled. This  is achieved via
a slow shearing procedure which has for an aim to probe each
static configuration by allowing the system to evolve at a very
slow shear rate. We first introduce a procedure to obtain a
dynamical measurement of the compactivity via a diffusion-mobility
protocol. We call this quantity the effective temperature
$T_{\mbox{\scriptsize eff}}$ and show that it satisfies a form of
the zero-th law of thermodynamics and thus has a thermodynamic
meaning.

The next crucial test for this assumption is to show that the
effective temperature obtained dynamically can also be obtained
via a flat average over the jammed configurations. Such a test has
been performed in \cite{nature}, where it was indeed shown that
$T_{\mbox{\scriptsize eff}}$ is very close to the compactivity of
the packing $X_E$. This result will be shown explicitly in Section
\ref{ergodicity}. We conclude that the jammed configurations
explored during shear are sampled in an equiprobable way as
required by the ergodic principle. Moreover the dynamical
measurement of compactivity renders the thermodynamic approach
amenable to experimental investigations.

In the next sections we calculate the effective temperature of the
packing dynamically and show its relation to the compactivity,
calculated employing a configurational average.

\subsection{Exploring the jammed configurations dynamically: effective
temperature $T_{\mbox{\scriptsize eff}}$} \label{teff}

Consider a `tracer' body of arbitrary shape immersed in a liquid
in thermal equilibrium. As a consequence of the irregular
bombardment by the particles of the surrounding liquid, the tracer
performs a diffusive, fluctuating `Brownian' motion.  The motion
is unbiased, and for large times the average square of the
displacement goes as $\langle |x(t)-x(0)|^2 \rangle = 2 D t $,
where $D$ is the diffusivity.  On the other hand, if we pull
gently on the tracer with a constant force $F$, the liquid
responds with a viscous, dissipative force.  The averaged
displacement after a large time is $\langle [x(t)-x(0)] \rangle =
F \chi t$, where $\chi$ is the mobility. Although both $D$ and
$\chi$ strongly depend on the shape and size of the tracer, they
turn out to always be related by the Einstein relation
\begin{equation}
T= D/\chi, \label{fdt}
\end{equation}
(a form of the Fluctuation-Dissipation Theorem, FDT), where $T$ is
the temperature of the liquid.

The Einstein relation
is strictly valid for equilibrium thermal systems. However, it has
been shown that fluctuation-dissipation relations are relevant to
describe the thermodynamics of out-of-equilibrium systems. Insight
into the validity of fluctuation-dissipation relations to describe
far from equilibrium systems first came from glass theory
\cite{virasoro,ktw}. Recent analytic schemes for glasses have
shown that a generalized FDT gives rise to a well-defined
effective temperature which is different from the bath temperature
\cite{ck}. It governs the heat flow and the slow components of
fluctuations and responses of all observables. Explicit
verifications of this approach have been made until present within
the mean-field/mode-coupling models of the glass transition
\cite{kurchan1}.
More recent studies have supported the existence of effective
temperatures in   schematic finite-dimensional models of glassy
systems \cite{nicodemi,bklm}.

The existence of a similar effective temperature for
out-of-equilibrium sheared granular systems is by no means
obvious. Moreover, its existence may provide a profound link
between the physics of granular materials and glasses. In the
absence of a first principle derivation, one needs to ascertain
its validity for every particular case experimentally, or
numerically at the least. In order to test the existence of an
effective temperature with a thermodynamic meaning for dense
slow-moving granular matter we perform a numerical study of a
diffusion-mobility computation in conditions that can be
reproduced in the laboratory \cite{nature}.

\begin{figure}
\centerline {\hbox{
\includegraphics*[width=.8\textwidth]{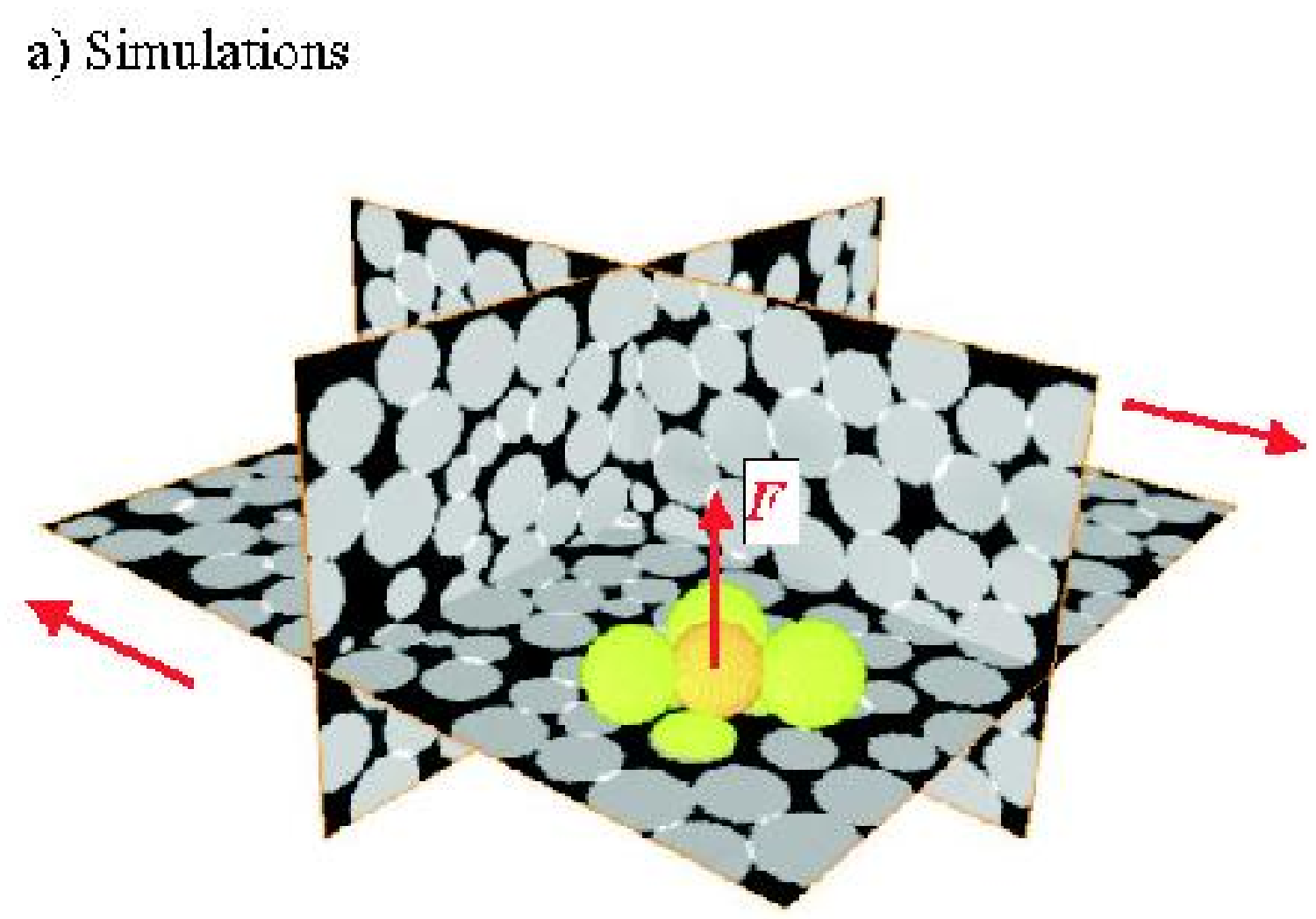} } }
\caption{Simulations of grains of 100 microns interacting via
Hertz-Mindlin contact forces. A slow shear flow, indicated by the
arrows, is applied to the jammed system. We follow the tracer
particle trajectories to obtain the diffusivity. An external force
$F$ is then applied to the tracers in response to which we measure
the particle mobility. These dynamical measurements yield an
"effective temperature" obtained from an Einstein relation which
is indeed very close to the compactivity obtained by a flat
average over the ensemble of jammed configurations.} \label{amira}
\end{figure}

We consider a system of large and small spherical grains in a
periodic cell. The simulations involve the application of a gentle
shear on the particles in the $y$-$z$ plane, at constant volume
(see Fig. \ref{amira}). The shearing mechanism moves the periodic
images at the top and the bottom of the cell with velocities $\dot
\gamma L/2$, where $\dot \gamma$ is the shear rate (Lees-Edwards
boundary conditions \cite{lees}). A linear, uniform velocity
profile along the $z-$direction  is obtained using a modified set
of equations to suit our computations (see Chapter 8 of
\cite{lees}). Periodic boundary conditions are enforced in the
$x-$direction and the $y-$direction of the flow. We focus our
study on the region of slow shear rates, where the system is
always close to jamming. We avoid shear bands by imposing a linear
velocity profile and avoid segregation which may occur at much
longer time scales than those employed in our computations.

The time resolved displacement of the particles are measured to
obtain their random fluctuations as well as the response function
of the tracers to allow the determination of the effective
temperatures via the Einstein relation (\ref{fdt}). We  measure
the spontaneous fluctuations $\langle |x(t) -x(0)|^2\rangle$ and
force-induced displacements $\langle [x(t)-x(0)]\rangle /F $,
where $F$ is a small external force in the $x$-direction, for two
types of tracers with different sizes. We calculate
$T_{\mbox{\scriptsize eff}}$ by using parametric plots of $\langle
|x(t)-x(0)|^2 \rangle$ versus $\langle [x(t)-x(0)] \rangle/F $
with
 $t$ as parameter.
The parametric procedure of obtaining the effective temperatures
 allows the study of
 $T_{\mbox{\scriptsize eff}}$ for different time scales \cite{ck,bklm}.
Hence we determine the temperature of the different modes of
relaxation (see also \cite{ohern3,liu-langer}).

\begin{figure}
\centerline {\hbox{
\includegraphics*[width=.67\textwidth]{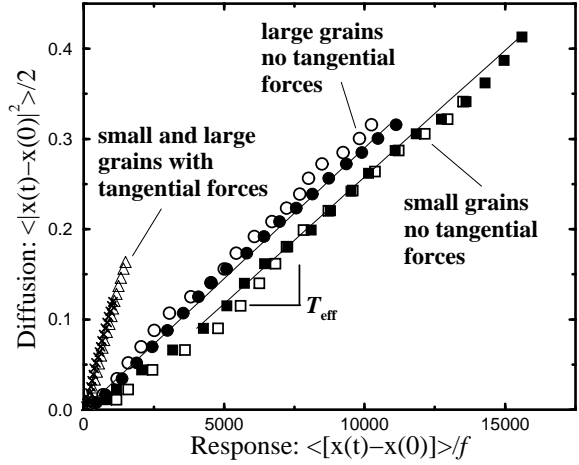} } }
\caption{Parametric plot of diffusion vs  response function  for
small and large grains and for spheres interacting with tangential
forces (grains) and without tangential forces (emulsions). The
fitting at long time scales shows the existence of a well-defined
temperature which is the same for small and large grains: $
T_{\mbox{\scriptsize eff}}=2.8 \times 10^{-5}$ for grains without
transverse forces and $ T_{\mbox{\scriptsize eff}}=1.2 \times
10^{-4}$ for grains with Mindlin transverse forces and Coulomb
friction. These effective temperatures (measured here in reduced
units \protect\cite{nature}) are in fact very large, i.e. $\sim
10^{14}$ times $k_BT$ at room temperature, as expected. We
calculate the response function for several small external fields
and find the same temperature indicating that we are in the linear
response regime. Plotted are results for a system  without
transverse forces using: $F=1.7 \times 10^{-5}$ (small grains
$\Box$, large grains $\circ$) and $F=2.6 \times 10^{-5}$ (small
grains $\protect\rule{2mm}{2mm}$, large grains $\bullet$). For a
system with tangential forces and Coulomb friction we show the
case $F=6 \times 10^{-5}$ (small grains $\bigtriangleup$, large
grains $\times$). } \label{ck}
\end{figure}

Our data is consistent with a granular Einstein relation (Fig.
\ref{ck}):
\begin{equation}
\langle |x(t) -x(0)|^2\rangle = 2 ~ T_{\mbox{\scriptsize eff}}
\frac{\langle [x(t)-x(0)]\rangle}{F}, \label{fdt2}
\end{equation}
valid for all tracers {\em with the same } fluctuation-dissipation
 temperature $T_{\mbox{\scriptsize eff}}$ for widely separated
time scales.

If there is an underlying thermodynamics, it will impose that the
effective temperature be independent of the tracer's  size: a
strong condition required for the thermodynamic hypothesis. We
then verify if all tracers have ``equilibrated'' at the same
effective temperature by calculating the effective temperature for
small and large tracers and finding that indeed they are the same.
Figure \ref{ck} shows the parametric plot of the diffusion versus
response function for systems with friction and frictionless
grains. In both cases we consider the temperature of the small and
large grains and obtain the same temperature. The fact that the
effective temperature is the same for both types of particles and
that it is only valid for the long-time relaxation suggest that
$T_{\mbox{\scriptsize eff}}$ can be considered to be the
temperature of the slow modes.

\subsection{Exploring the jammed configurations via a flat average.
Test of ergodicity: $T_{\mbox{\scriptsize eff}} = X_E$}
\label{ergodicity}

The systems under investigation have exponentially large (in the
number of particles) number of stable states jammed at zero bath
temperature. In the previous section we explored such an energy
landscape via slow shear. Next, we  develop an independent method
to study the configurational space. It  allows us to investigate
the statistical properties of the jammed states available at a
given energy and volume. In turn we  investigate whether it is
possible to relate the dynamical temperature obtained above via a
diffusion-mobility protocol to the configurational compactivity
based on jammed states.

In order to calculate $X_{\mbox{\scriptsize E}}$ and compare with
the obtained $T_{\mbox{\scriptsize eff}}$ we need to sample the
jammed configurations at a given energy and volume in a {\it
equiprobable} way. In order to do this we sample the jammed
configurations with the following probability distribution:
\begin{equation}
P_\nu \sim \exp[- E^\nu/T^* - E^\nu_{\mbox{\scriptsize
jammed}}/T_{\mbox{\scriptsize aux}}] \label{partition}
\end{equation}
Here the deformation energy $E$ corresponds to the  Hertzian
energy of deformation of the grains or the Princen energy for
droplets. The extra term added in Eq. (\ref{partition})
 allows us to perform the flat sampling
of the jammed states and plays the role of the term $\Theta_\nu$
in Eq. (\ref{canonical2}). The jammed energy is such that it
vanishes at the jammed configurations:
\begin{equation}
E_{\mbox{\scriptsize jammed}} \propto \sum_a
\left|\vec{F}_a\right|^2,
\end{equation}
where $\vec{F}_a$ is the total force exerted on particle $a$ by
its neighbours.

\begin{figure}
\centerline {\hbox{
\includegraphics*[width=.67\textwidth]{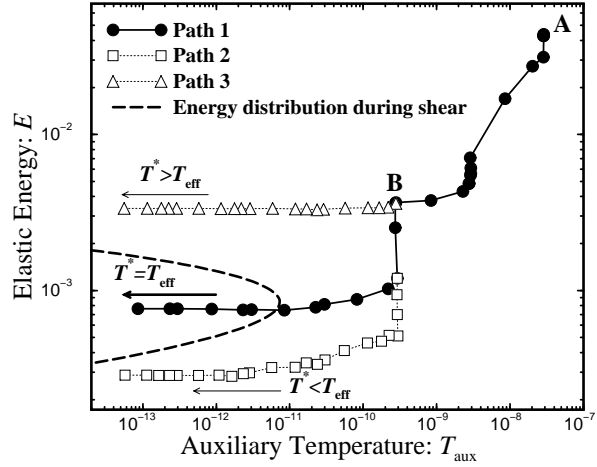} } }
\caption{ Annealing procedure to calculate  $X_E$ at different
elastic compressional energies.  We plot the elastic energy vs
$T_{\mbox{\scriptsize aux}}$ during  annealing together with the
distribution of elastic energies obtained during shear (dashed
curve, mean value $\langle E \rangle = 8.4\times 10^{-4}$). We
equilibrate the system for $40\times 10^6$ iterations at A:
$(T^*=3.4\times 10^{-2}, T_{\mbox{\scriptsize aux}}=3\times
10^{-8})$. We then anneal slowly both temperatures until B:
$(T^*=3.4\times 10^{-4}, T_{\mbox{\scriptsize aux}}=3\times
10^{-10})$, where we split the trajectory in three paths in the
$(T^*, T_{\mbox{\scriptsize aux}})$ plane.  Path 1: we anneal $
T_{\mbox{\scriptsize aux}}\to 0$ and $T^* \to 2.8\times 10^{-5}$
which corresponds to $ T_{\mbox{\scriptsize eff}}$ obtained during
shear (Fig. \protect\ref{ck}).  Path 2: we anneal $
T_{\mbox{\scriptsize aux}}\to 0$ and $T^* \to 3.4\times 10^{-6}$.
Path 3: we anneal $ T_{\mbox{\scriptsize aux}}\to 0$ but keep
$T^*=3.4\times 10^{-4}$ constant.  When we set $T^* =
T_{\mbox{\scriptsize eff}}$ (Path 1), the final elastic
compressional energy value when $ T_{\mbox{\scriptsize aux}}\to 0$
falls inside the distribution of energies obtained, and it is very
close to the mean value of the elastic energy during shear
$\langle E \rangle$. This proves that $ T_{\mbox{\scriptsize eff}}
= X_E$ under the numerical accuracy of the simulations.  For other
values of $T^* \ne T_{\mbox{\scriptsize eff}}$ the final $E$ falls
out of the distribution obtained during shear (Paths 2 and 3).  We
also follow different trajectories (not shown in the figure) to
$T^* \to 2.8\times 10^{-5}, T_{\mbox{\scriptsize aux}}\to 0$ and
find the same results indicating that our procedure is independent
of the annealing path. } \label{annealing}
\end{figure}

We introduce two ``bath'' temperatures (these temperatures are
$\sim 10^{14}$ times the room temperature) which will allow us to
explore the configuration space and calculate the entropy of the
packing assuming a flat average over the jammed configurations. We
perform equilibrium MD simulations with two auxiliary ``bath''
temperatures $(T^*,T_{\mbox{\scriptsize aux}})$, corresponding to
the partition function (\ref{partition}). Annealing
$T_{\mbox{\scriptsize aux}}$ to zero selects the jammed
configurations ($E_{\mbox{\scriptsize jammed}}=0$), while $T^*$
fixes the energy $E$.

In practice we perform equilibrium MD simulations with a modified
potential energy:

\begin{equation}
U = \frac{T_{\mbox{\scriptsize aux}}}{T^*} E +
E_{\mbox{\scriptsize jammed}},
\end{equation}
and calculate the force on each particle from $\vec{F} = -
\vec{\nabla} U$. Since we need to calculate the force from a
potential energy, only conservative systems can be studied with
this method. Thus we focus our calculations on the system of
frictionless particles. The auxiliary temperature
$T_{\mbox{\scriptsize aux}}$ is controlled by a thermostat which
adjusts the velocities of the particles to a kinetic energy
determined by $T_{\mbox{\scriptsize aux}}$. We start by
equilibrating the system at high temperatures
($T_{\mbox{\scriptsize aux}}$ and $T^*$ $\sim \infty$) and anneal
slowly the value $T_{\mbox{\scriptsize aux}}$ to zero and tune
$T^*$ so as to reach the  value of $E$ that corresponds to the
average deformation energy  obtained during shear.

The partition function is
\begin{equation}
Z = \sum_\nu \exp[- E^\nu/T^* - E^{\nu}_{\mbox{\scriptsize
jammed}}/T_{\mbox{\scriptsize aux}}], \label{partition2}
\end{equation}
from where we obtain the compactivity as
\begin{equation}
T^* = \frac{\partial E}{\partial S} \stackrel {\mbox{\scriptsize
$T_{\mbox{\scriptsize aux}}\to 0$}}{\longrightarrow} X_E,
\end{equation}
Thus at the end of the annealing process ($T_{\mbox{\scriptsize
aux}}\to 0$), $T^*(E)=X_{E}(E)$, since in this limit we are
sampling the configurations with vanishing fraction of moving
particles at a given $E$.

At the end of the protocol the compactivity at a given deformation
energy can be obtained as illustrated in Fig. \ref{annealing}. The
remarkably result is that  the compactivity and the effective
temperature obtained dynamically are found to coincide to within
the computational error \cite{nature},
\begin{equation}
X_E \approx T_{\mbox{\scriptsize eff}}.
\end{equation}

This provides strong evidence for the validity of the effective
temperature as a dynamical estimate of the compactivity, and more
importantly, justifies the use of the novel statistical
measurements we have presented in fully characterising the
macroscopic properties of the system.

To summarise,  the fact that slow relaxation modes can be
characterized by a temperature raises the question of the
existence of a form of ergodicity for the structural motion,
allowing a construction of a statistical mechanics ensemble for
the slow motion of the grains. This argument leads us back to the
ideas of the thermodynamics of jammed states. In parallel to these
dynamical measurements, the same information is drawn from the
system by a flat statistical average over the jammed
configurations. Once all the static configurations have been
visited by the system, the compactivity $X_E$ can be calculated
from the statistics of the canonical ensemble of the jammed
states. The logarithm of the available configurations at a given
energy and volume reveals the entropy, from which the compactivity
is calculated. Our explicit computation shows that the temperature
arising from the Einstein relation (\ref{fdt2}) can be understood
in terms of the configurational compactivity $X_{E}$ arising from
the statistical ensemble of jammed states. This provides a strong
evidence for the validity of the thermodynamic approach.

\printindex
\end{document}